\pdfoutput=1
\documentclass[useAMS,usenatbib]{mn2e}
\usepackage{apjfonts}
\usepackage{graphicx}
\usepackage{amsmath}
\usepackage{amssymb}
\usepackage{ctable}
\usepackage{fixltx2e}
\usepackage{threeparttable}
\usepackage{hyperref}
\hypersetup{colorlinks=true,linkcolor=blue,citecolor=blue,filecolor=blue,urlcolor=blue}

\newcommand{\be}{\begin{equation}}
\newcommand{\ee}{\end{equation}}
\newcommand{\ba}{\begin{eqnarray}}
\newcommand{\ea}{\end{eqnarray}}
\newcommand{\Mvir}{M_{\rm vir}}
\newcommand{\Rvir}{R_{\rm vir}}

\newcommand{\Mhalo}{M_{\rm halo}}

\newcommand{\Ms}{M_{\ast}}

\newcommand{\cm}{{\rm cm}}

\newcommand{\pc}{{\rm pc}}
\newcommand{\kpc}{{\rm kpc}}
\newcommand{\Mpc}{{\rm Mpc}}

\newcommand{\s}{{\rm s}}

\newcommand{\g}{{\rm g}}

\newcommand{\Msun}{M_{\sun}}

\newcommand{\Zsun}{Z_{\sun}}

\newcommand{\HII}{H\,{\sc ii}}

\newcommand{\mb}{m_b}
\newcommand{\mDM}{m_{\rm DM}}
\newcommand{\eg}{\epsilon_{\rm gas}}
\newcommand{\es}{\epsilon_{\rm star}}
\newcommand{\eDM}{\epsilon_{\rm DM}}
\newcommand{\hinv}{h^{-1}}
\newcommand{\nc}{n_{\rm th}}

\newcommand{\fesc}{f_{\rm esc}}
\newcommand{\fave}{\langle\fesc\rangle}
\newcommand{\fage}{\fave_{\rm age}}

\newcommand{\dd}{{\rm d}}
\newcommand{\LCDM}{$\Lambda$CDM}
\newcommand{\aj}{AJ}
\newcommand{\apj}{ApJ}
\newcommand{\apjl}{ApJ}
\newcommand{\apjs}{ApJS}
\newcommand{\mnras}{MNRAS}
\newcommand{\aap}{A\&A}
\newcommand{\araa}{ARA\&A}
\newcommand{\nat}{Nature}
\newcommand{\pasj}{PASJ}
\newcommand{\pasa}{PASA}
\newcommand{\rmxaa}{RMxAA}

\title[$\fesc$ in FIRE-2 simulations]
{No missing photons for reionization: moderate ionizing photon escape fractions from the FIRE-2 simulations}

\author[X. Ma et al.]{
  \parbox[t]{1.0\textwidth}{
   Xiangcheng Ma,$^{1}$\thanks{E-mail: \href{mailto:xchma@berkeley.edu}{xchma@berkeley.edu}}
   Eliot Quataert,$^1$
   Andrew Wetzel,$^2$
   Philip F. Hopkins,$^3$ \\
   Claude-Andr{\'e} Faucher-Gigu{\`e}re,$^4$
   Du{\v s}an Kere{\v s}$^5$
  }
  \vspace{5pt} \\
  $^1$Department of Astronomy and Theoretical Astrophysics Center, University of California Berkeley, Berkeley, CA 94720 \\
  $^2$Department of Physics, University of California, Davis, CA 95616, USA \\
  $^3$TAPIR, MC 350-17, California Institute of Technology, Pasadena, CA 91125, USA \\ 
  $^4$Department of Physics and Astronomy and CIERA, Northwestern University, 2145 Sheridan Road, Evanston, IL 60208, USA \\
  $^5$Department of Physics, Center for Astrophysics and Space Sciences, University of California at San Diego, 9500 Gilman Drive, La Jolla, CA 92093
}

\pagerange{\pageref{firstpage}--\pageref{lastpage}}
\date{Draft version \today}

\begin{document}
\maketitle
\label{firstpage}

\begin{abstract}
We present the escape fraction of hydrogen ionizing photons ($\fesc$) from a sample of 34 high-resolution cosmological zoom-in simulations of galaxies at $z\geq5$ in the Feedback in Realistic Environments project, post-processed with a Monte Carlo radiative transfer code for ionizing radiation. Our sample consists of 8500 halos in $\Mvir\sim10^8$--$10^{12}\,\Msun$ ($\Ms\sim10^4$--$10^{10}\,\Msun$) at $z=5$--12. We find the sample average $\fave$ increases with halo mass for $\Mvir\sim10^8$--$10^{9.5}\,\Msun$, becomes nearly constant for $10^{9.5}$--$10^{11}\,\Msun$, and decreases at $\gtrsim10^{11}\,\Msun$. Equivalently, $\fave$ increases with stellar mass up to $\Ms\sim10^8\,\Msun$ and decreases at higher masses. Even applying single-star stellar population synthesis models, we find a moderate $\fave\sim0.2$ for galaxies at $\Ms\sim10^8\,\Msun$. Nearly half of the escaped ionizing photons come from stars 1--3\,Myr old and the rest from stars 3--10\,Myr old. Binaries only have a modest effect, boosting $\fave$ by $\sim$25--35\% and the number of escaped photons by 60--80\%. Most leaked ionizing photons are from vigorously star-forming regions that usually contain a feedback-driven kpc-scale superbubble surrounded by a dense shell. The shell is forming stars while accelerated, so new stars formed earlier in the shell are already inside the shell. Young stars in the bubble and near the edge of the shell can fully ionize some low-column-density paths pre-cleared by feedback, allowing a large fraction of their ionizing photons to escape. The decrease of $\fave$ at the high-mass end is due to dust attenuation, while at the low-mass end, $\fave$ decreases owing to inefficient star \linebreak formation and hence feedback. At fixed mass, $\fave$ tends to increase with redshift. Although the absolute $\fave$ does not fully converge with resolution in our simulations, the mass- and redshift-dependence of $\fave$ is likely robust. Our simulations produce sufficient ionizing photons for cosmic reionization.
\end{abstract}

\begin{keywords}
galaxies: evolution -- galaxies: formation -- galaxies: high-redshift -- cosmology: theory -- dark ages, reionization, first stars
\vspace{-5pt}
\end{keywords}

\section{Introduction}
\label{sec:intro}
Over the past decade, thanks to a series of deep imaging campaigns carried out with the {\em Hubble Space Telescope} (HST) and ground-based facilities, we have obtained relatively robust constraints on the bright-end ($\rm M_{UV}\lesssim-17$) rest-frame ultraviolet luminosity functions (UVLFs) of star-forming galaxies up to $z\sim8$ \citep[e.g.][]{Bouwens:2015,Bowler:2015,Finkelstein:2015,Ono:2018} and tentative measurements on the UVLFs at $z\sim9$--10 \citep[e.g.][]{Oesch:2013,Oesch:2014,Bouwens:2016a,Stefanon:2017a}. The Hubble Frontier Fields (HFFs) campaign make it even possible to probe the UVLFs down to $\rm M_{UV}\lesssim-12$ at $z\sim6$ \citep[e.g.][]{Atek:2015a,Atek:2018a,Bouwens:2017a,Livermore:2017}. The upcoming {\em James Webb Space Telescope} (JWST; scheduled launch date in March 2021) is expected to remarkably advance our knowledge on the galaxy populations at these redshifts.

These high-redshift star-forming galaxies are thought to be the dominant sources for reionization (e.g. \citealt{Madau:1999,Faucher-Giguere:2009,Haardt:2012,Faucher-Giguere:2020}; however, see \citealt{Madau:2015}), a phase transition of the hydrogen in the intergalactic medium (IGM) from neutral to fully ionized. Cosmic reionization began with the onset of the first generation of stars at $z\sim20$--30 \citep[e.g.][]{Loeb:2001} and finished by $z\sim5$ \citep[e.g.][]{Fan:2006,Becker:2015}. The number of ionizing photons emitted from high-redshift galaxies per unit time can be estimated from the observed UVLFs and stellar population synthesis models \citep[e.g.][]{Leitherer:1999,Conroy:2013,Eldridge:2017}. One critical, yet poorly understood, parameter to link high-redshift galaxies to cosmic reionization is the escape fraction of ionizing photons from these galaxies to the IGM ($\fesc$).

Models that describe the reionization history from the galaxy populations at $z\geq5$ need to make assumptions about $\fesc$, either a constant $\fesc$ for all galaxies or some mass- and redshift-dependent form of $\fesc$ \citep[e.g.][]{Finkelstein:2012,Finkelstein:2019,Kuhlen:2012,Robertson:2013,Robertson:2015,Bouwens:2015a,Naidu:2020,Yung:2020}. Currently available observational constraints on the reionization history, such as the integrated Thomson scattering optical depths \citep[e.g.][]{Planck-Collaboration:2016,Planck-Collaboration:2018}, Lyman-$\alpha$ transmission in the Gunn--Peterson trough \citep[e.g.][]{Fan:2006a,Becker:2015}, the fraction of Lyman-$\alpha$ \linebreak emitters \citep[e.g.][]{Stark:2011,Mesinger:2015}, and the distribution of Lyman-$\alpha$ equivalent widths \citep[EWs; e.g.][]{Mason:2018,Mason:2019}, suggest $\fesc\sim0.2$, although a robust constraint on $\fesc$ and its mass and redshift dependence, cannot yet be achieved.

Direct detection of ionizing or Lyman-continuum (LyC) fluxes from galaxies at $z\gtrsim5$ is almost impossible. Great efforts have been made to search for rest-frame LyC fluxes from galaxies at $z\sim0$--4 over the past two decades. \citet{Steidel:2001} reported detection of strong LyC flux and inferred high $\fesc$\footnote{Note that \citet{Steidel:2001} measured the {\em relative} escape fraction $f_{\rm esc}^{\rm rel}\gtrsim0.5$, defined as $\fesc$ divided by the escape fraction of {1500\,\AA} photons.} from a stacked spectrum of 29 galaxies at $\langle z\rangle\sim3.4$ \citep[see also][]{Shapley:2006}, but these early studies at $z\gtrsim3$ likely suffer from foreground contaminations from low-redshift interlopers \citep[e.g.][]{Vanzella:2010,Siana:2015}. In the literature, various authors have reported significantly lower $\fesc$ (of order 0.01) using galaxy samples from the local Universe to $z\sim3$ \citep[e.g.][]{Leitherer:1995,Cowie:2009,Bridge:2010,Siana:2010,Boutsia:2011,Leitet:2011,Leitet:2013,Grazian:2016,Grazian:2017,Rutkowski:2016}. Such low $\fesc$ is not sufficient for cosmic reionization.

The field has turned around in recent years. Strong LyC leakage has been confirmed from a number of galaxies at $z\sim0$--4, with inferred $\fesc$ from a few per\,cent to over 50 per\,cent \citep[e.g.][]{Vanzella:2012,Vanzella:2016,Vanzella:2020,Izotov:2016a,Izotov:2016,Izotov:2018,Fletcher:2019,Rivera-Thorsen:2019,Ji:2020}. Most of them are compact, extreme starburst galaxies that are thought to be analogs of typical star-forming galaxies in the reionization era. The strong LyC-leaking galaxies share some common properties, such as high \linebreak $\rm [O\,{\textsc{iii}}]/[O\,{\textsc{ii}}]$ ratios \citep[O32; e.g.][]{Nakajima:2014,Izotov:2016}, high Lyman-$\alpha$ escape fractions and EWs, double-peak Lyman-$\alpha$ profile with small peak separations \citep[e.g.][]{Verhamme:2015,Verhamme:2017}, and weak low-ionization metal absorption lines \citep[e.g.][]{Jaskot:2014,Chisholm:2018a}. There also exist galaxies with comparable O32 ratios and Lyman-$\alpha$ features that do not have \linebreak detectable LyC fluxes, which are thought to be line-of-sight variations \citep[e.g.][]{Jaskot:2019,Malkan:2019,Nakajima:2020,Izotov:2020}. Moderate $\fesc\sim0.1$--0.2 have also been reported recently for considerably large spectroscopic samples of galaxies at $z\sim3$ after a careful examination of foreground contamination \citep[e.g.][]{Nestor:2013,Steidel:2018}. Newly developed cross correlation analysis between star-forming galaxies and IGM transmission at $z\geq5$ also suggests $\fesc\sim0.1$--0.2 \citep[e.g.][]{Kakiichi:2018,Meyer:2019}.

In principle, the common features of strong LyC leakers outlined above may be used as an indirect indicator of $\fesc$ from high-redshift galaxies once rest-frame UV-to-optical spectra are accessible at $z\geq5$ with JWST, potentially offering an independent probe of the contribution of high-redshift galaxies to cosmic reionization. The most important question now is to understand the key physics that governs the escape of ionizing photons, which is also a critical prerequisite for understanding those possible indirect indicators of $\fesc$. High-resolution, spatially resolved spectroscopic data of some LyC leakers in the nearby Universe have been obtained recently to address this question \citep[e.g.][]{Keenan:2017,Micheva:2018,Menacho:2019}. A comparably detailed theoretical investigation is demanded by these observations.

The escape of ionizing photons involves physics spanning several orders of magnitude in scale. The young stars that produce the majority of the ionizing photons are normally surrounded by dense gas left from the giant molecular clouds (GMCs) in which the stars are formed. Most ionizing photons from the young stars will be absorbed locally before the birth clouds are dispersed \citep[e.g.][]{Kim:2013,Ma:2015,Kimm:2017,Kakiichi:2019,Kim:2019}. The time-scale for cloud destruction has to compete with time-scale of massive star evolution, as the ionizing photon budget of a stellar population declines rapidly after the death of the most massive stars in about 3\,Myr. Some ionizing photons may also be absorbed by the extended neutral hydrogen in the interstellar medium (ISM) and in the halo before the rest photons escape to the IGM \citep[e.g.][]{Ferrara:2013}. Given the complexity, hydrodynamic simulations of galaxy formation form a powerful tool for understanding the escape of ionizing photons.

Early studies using (sub-)kpc-resolution cosmological simulations that cannot properly resolve the ISM structure tend to produce high $\fesc$ from tens of per\,cent to unity (e.g. \citealt{Razoumov:2010,Yajima:2011}; see also \citealt{Anderson:2017}). Intriguingly, with more detailed treatments of the multi-phase ISM and feedback developed for newer simulations of better resolution, in which the formation and feedback destruction of GMCs start to be resolved, the predicted $\fesc$ has been brought down significantly to a few to ten per\,cent (at least in halos above $\Mvir\sim10^8\,\Msun$; e.g. \citealt{Gnedin:2008,Paardekooper:2011,Paardekooper:2015,Kim:2013,Kimm:2014,Wise:2014,Ma:2015,Ma:2016,Xu:2016b,Rosdahl:2018}; however, \citealt{Wise:2009}). \citet{Ma:2015} argued in their simulations, ionizing photons from stars younger than 3\,Myr are almost entirely consumed by the birth clouds, while there is no longer a sufficiently large ionizing photon budget available from older stars, thereby resulting in $\fesc\lesssim0.05$. \linebreak Such $\fesc$ is at the lower limit of what required for cosmic reionization, hence runaway OB stars \citep[e.g.][]{Conroy:2012,Kimm:2014} and binaries \citep[e.g.][]{Ma:2016,Rosdahl:2018} are invoked to provide the `missing' photons, either by making the \linebreak ionizing photons from young stars escape more easily or producing more ionizing photons after 3\,Myr.

\begin{table*}
\caption{A list of simulations used in this paper.}
\begin{threeparttable}
\begin{tabular}{cccccccc|cccccccc}
\hline
Name & $z_{\rm f}$ & $\Mhalo$ & $\Ms$ & $\mb$ & $\mDM$ & $\eg$ & $\eDM$ & Name & $z_{\rm f}$ & $\Mhalo$ & $\Ms$ & $\mb$ & $\mDM$ & $\eg$ & $\eDM$ \\
 & & [$\Msun$] & [$\Msun$] & [$\Msun$]  & [$\Msun$] & [pc] & [pc] & & & [$\Msun$] & [$\Msun$] & [$\Msun$]  & [$\Msun$] & [pc] & [pc] \\
\hline
z5m12b & 5 & 8.73e11 & 2.55e10 & 7126.5 & 3.9e4 & 0.42 & 42 & z5m10c & 5 & 1.34e10 & 5.58e7 & 954.4 & 5.2e3 & 0.28 & 21 \\
z5m12c & 5 & 7.91e11 & 1.83e10 & 7126.5 & 3.9e4 & 0.42 & 42 & z5m10b & 5 & 1.25e10 & 3.42e7 & 954.4 & 5.2e3 & 0.28 & 21 \\
z5m12d & 5 & 5.73e11 & 1.20e10 & 7126.5 & 3.9e4 & 0.42 & 42 & z5m10a & 5 & 6.60e9 & 1.48e7 & 119.3 & 650.0 & 0.14 & 10 \\
z5m12e & 5 & 5.04e11 & 1.35e10 & 7126.5 & 3.9e4 & 0.42 & 42 & z5m09b & 5 & 3.88e9 & 2.79e6 & 119.3 & 650.0 & 0.14 & 10 \\
z5m12a & 5 & 4.51e11 & 5.36e9 & 7126.5 & 3.9e4 & 0.42 & 42 & z5m09a & 5 & 2.36e9 & 1.64e6 & 119.3 & 650.0 & 0.14 & 10 \\
z5m11f & 5 & 3.15e11 & 4.68e9 & 7126.5 & 3.9e4 & 0.42 & 42 & z7m12a & 7 & 8.91e11 & 1.66e10 & 7126.5 & 3.9e4 & 0.42 & 42 \\
z5m11e & 5 & 2.47e11 & 2.53e9 & 7126.5 & 3.9e4 & 0.42 & 42 & z7m12b & 7 & 6.40e11 & 1.44e10 & 7126.5 & 3.9e4 & 0.42 & 42 \\
z5m11g & 5 & 1.98e11 & 1.86e9 & 7126.5 & 3.9e4 & 0.42 & 42 & z7m12c & 7 & 4.71e11 & 1.16e10 & 7126.5 & 3.9e4 & 0.42 & 42 \\
z5m11d & 5 & 1.35e11 & 1.62e9 & 7126.5 & 3.9e4 & 0.42 & 42 & z7m11a & 7 & 3.32e11 & 7.17e9 & 7126.5 & 3.9e4 & 0.42 & 42 \\
z5m11h & 5 & 1.01e11 & 1.64e9 & 7126.5 & 3.9e4 & 0.42 & 42 & z7m11b & 7 & 2.48e11 & 2.00e9 & 7126.5 & 3.9e4 & 0.42 & 42 \\
z5m11c & 5 & 7.57e10 & 9.45e8 & 890.8 & 4.9e3 & 0.28 & 21 & z7m11c & 7 & 1.63e11 & 1.81e9 & 7126.5 & 3.9e4 & 0.42 & 42 \\
z5m11i & 5 & 5.17e10 & 2.77e8 & 890.8 & 4.9e3 & 0.28 & 21 & z9m12a & 9 & 4.20e11 & 1.24e10 & 7126.5 & 3.9e4 & 0.42 & 42 \\
z5m11b & 5 & 4.02e10 & 1.67e8 & 890.8 & 4.9e3 & 0.28 & 21 & z9m11a & 9 & 2.88e11 & 3.46e9 & 7126.5 & 3.9e4 & 0.42 & 42 \\
z5m11a & 5 & 4.16e10 & 1.22e8 & 954.4 & 5.2e3 & 0.28 & 21 & z9m11b & 9 & 2.23e11 & 3.49e9 & 7126.5 & 3.9e4 & 0.42 & 42 \\
z5m10f & 5 & 3.30e10 & 1.56e8 & 954.4 & 5.2e3 & 0.28 & 21 & z9m11c & 9 & 1.76e11 & 2.41e9 & 7126.5 & 3.9e4 & 0.42 & 42 \\
z5m10e & 5 & 2.57e10 & 3.93e7 & 954.4 & 5.2e3 & 0.28 & 21 & z9m11d & 9 & 1.28e11 & 1.46e9 & 7126.5 & 3.9e4 & 0.42 & 42 \\
z5m10d & 5 & 1.87e10 & 4.81e7 & 954.4 & 5.2e3 & 0.28 & 21 & z9m11e & 9 & 1.16e11 & 1.49e9 & 7126.5 & 3.9e4 & 0.42 & 42 \\
\hline
\end{tabular}
\begin{tablenotes}
\item Parameters describing the initial conditions and final galaxy properties of our simulations:
\item (1) $z_{\rm f}$: The redshift which the zoom-in region is selected at and the simulation is run to.
\item (2) $\Mhalo$ and $\Ms$: Halo mass and total stellar mass within the halo virial radius of the central halo at $z_{\rm f}$.
\item (3) $\mb$ and $\mDM$: Initial baryonic and DM particle mass in the high-resolution region. The masses of DM particles are fixed throughout the simulation. The masses of baryonic (gas and stars) particles are allowed to vary within a factor of two owing to mass loss and mass return due to stellar evolution.
\item (4) $\eg$ and $\eDM$: Plummer-equivalent force softening lengths for gas and DM particles, in comoving units above $z=9$ and physical units thereafter. Force softening for gas is adaptive ($\eg$ is the minimum softening length). Force softening length for star particles is $\es=5\eg$.
\end{tablenotes}
\end{threeparttable}
\label{tbl:sims}
\end{table*}

More importantly, this suggests that the `sub-grid' models implemented in these simulations have a large impact on the prediction of $\fesc$. That being said, this problem should be revisited while simulations are advancing in resolution and sub-grid recipes. Current state-of-the-art cosmological simulations of $z\geq5$ galaxies are fairly successful in reproducing the observed UVLFs at these redshifts \citep[e.g.][]{Gnedin:2016,Ocvirk:2016,Ceverino:2017,Ma:2018a,Ma:2019,Rosdahl:2018,Wilkins:2018}. It is worth emphasizing that large-volume simulations with (sub-)kpc resolution are not well suited for studying $\fesc$. Sufficiently detailed treatments of the ISM and feedback physics are mandatory. In this paper, we use a suite of 34 high-resolution cosmological zoom-in simulations of $z\geq5$ galaxies from the Feedback in Realistic Environments project \citep[FIRE;][]{Hopkins:2018b}.\footnote{The FIRE project website is at \url{https://fire.northwestern.edu}.} These simulations use the FIRE-2 version of the source code {\sc gizmo}\footnote{\url{http://www.tapir.caltech.edu/~phopkins/Site/GIZMO.html}} \citep{Hopkins:2015} that includes an explicit treatment of the multi-phase ISM, star formation, and stellar feedback at the smallest resolved scale. 

FIRE-2 is an updated version of the FIRE-1 simulations from \citet{Hopkins:2014}, including a newly developed hydrodynamic method, a more accurate implementation for mechanical feedback from supernovae (SNe) and stellar winds in \citet{Hopkins:2018a}, and more subtle differences \citep[see][for details]{Hopkins:2018b}. In this paper, we present a major update to our previous studies on $\fesc$ using the FIRE-1 simulations from \citet[][]{Ma:2015,Ma:2016}. The FIRE simulations are shown to reproduce a broad range of observed galaxy properties at $z\sim0$--6 \citep[][and references therein]{Hopkins:2018b}. In particular, the simulations studied in this paper produce excellent agreement with the observed galaxy UVLFs at $z\geq5$ \citep{Ma:2018a,Ma:2019}.

We post-process these simulations using a three-dimensional Monte Carlo radiative transfer (MCRT) code for ionizing radiation to calculate $\fesc$ from 8500 relatively well-resolved galaxies at $z\sim5$--12 spanning halo masses $\Mvir\sim10^8$--$10^{12}\,\Msun$ and stellar masses $\Ms\sim10^4$--$10^{10}\,\Msun$. We also investigate the dependence of $\fesc$ on galaxy mass and redshift in our simulated sample. Our results provide an essential complement to previous studies from other groups that found a decreasing $\fesc$ with halo mass in $\Mvir\sim10^6$--$10^9\,\Msun$ \citep[e.g.][]{Wise:2014,Paardekooper:2015,Xu:2016b} by extending the mass-dependence of $\fesc$ to higher masses. It is also \linebreak a key ingredient for modeling the reionization history as reviewed above \citep[e.g.][]{Finkelstein:2019,Yung:2020}.

We caution that the prediction of $\fesc$ is sensitive to the resolution and sub-grid models adopted in our simulations. Therefore, in this paper, we will mainly focus on the {\em qualitative} behaviors rather than the quantitative details of $\fesc$. We argue that this caveat may also apply to other studies on $\fesc$ using hydrodynamic simulations. In fact, our results of $\fesc$ do not yet fully converge with resolution. In Section \ref{sec:sims}, we describe our simulation sample and the ISM, star formation, and stellar feedback model adopted in our simulations. We introduce the MCRT code for our post-processing calculations in Section \ref{sec:mcrt}. Section \ref{sec:fesc} presents $\fesc$ for our simulations, the mass- and redshift-dependence of $\fesc$, and the effects of dust attenuation and binary stars on $\fesc$. In Section \ref{sec:physics}, we investigate the most critical physics that governs the escape of ionizing photons. We discuss \linebreak our results in Section \ref{sec:discussion} and conclude in Section \ref{sec:conclusion}. Throughout this paper, we define $\fesc$ of a galaxy as the absolute fraction of ionizing photons that escape the halo virial radius.

We adopt a standard flat {\LCDM} cosmology with {\it Planck} 2015 cosmological parameters $H_0=68 {\rm\,km\,s^{-1}\,Mpc^{-1}}$, $\Omega_{\Lambda}=0.69$, $\Omega_{m}=1-\Omega_{\Lambda}=0.31$, $\Omega_b=0.048$, $\sigma_8=0.82$, and $n=0.97$ \citep{Planck-Collaboration:2016a}. We use a \citet{Kroupa:2002} initial mass function (IMF) from 0.1--$100\,\Msun$, with IMF slopes of $-1.30$ from 0.1--$0.5\,\Msun$ and $-2.35$ from 0.5--$100\,\Msun$.

\begin{figure}
\centering
\includegraphics[width=\linewidth]{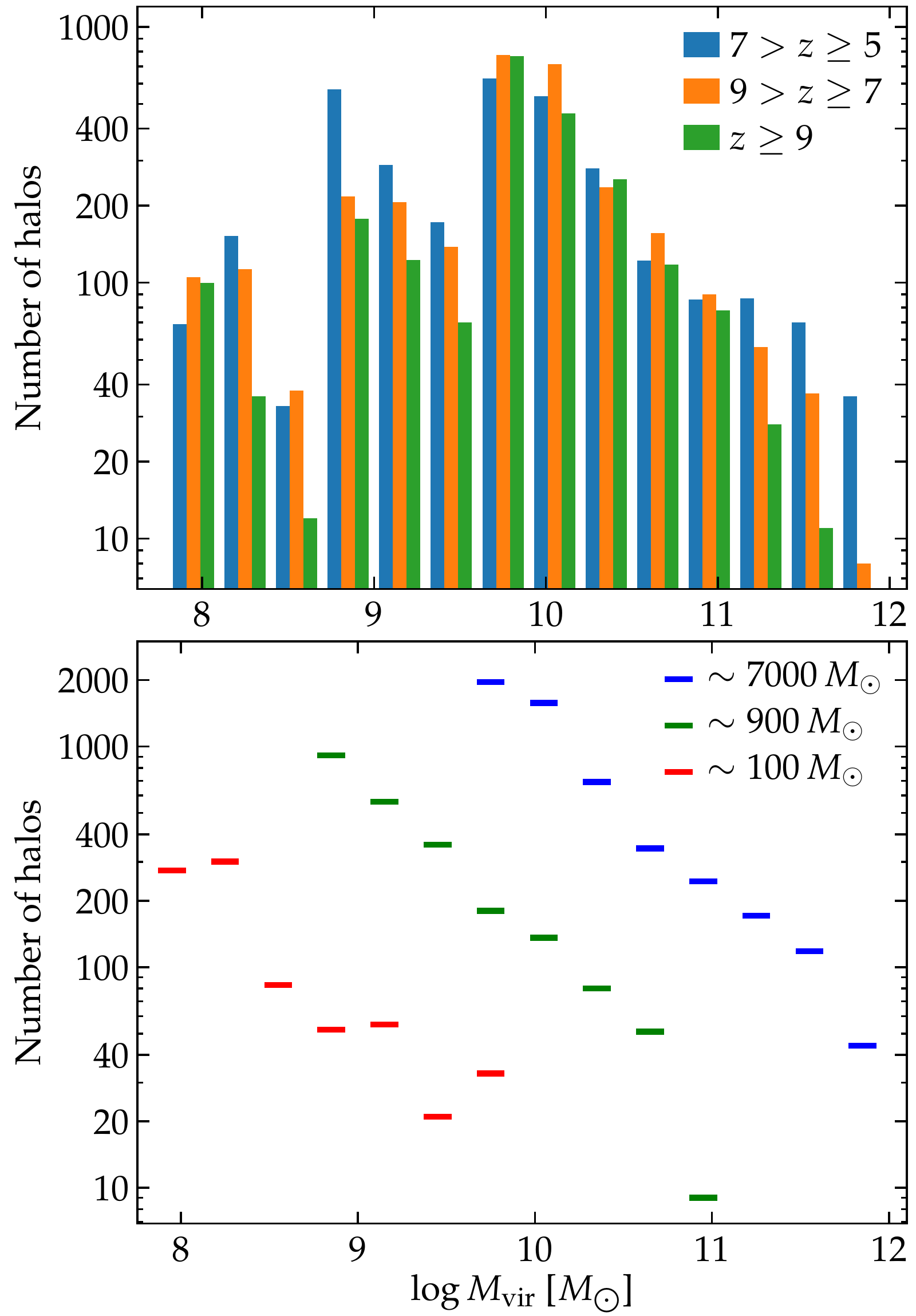}
\caption{Number of halo snapshots for every 0.3\,dex from $\Mhalo=10^{7.8}$--$10^{12}\,\Msun$ in our simulated sample. In the top panel, we divide our sample into three redshift bins regardless of mass resolution. In the bottom panel, we show the sample size at three mass resolution but at all redshift.}
\label{fig:sample} 
\end{figure}

\section{Method}
\label{sec:method}

\subsection{The simulations}
\label{sec:sims}
This work uses a suite of 34 cosmological zoom-in simulations at $z\geq5$, which we summarize in Table \ref{tbl:sims}. The sample is nearly identical to that presented in \citet{Ma:2019}, except that simulations z5m10a, z5m11c, and z5m11i are re-run from the same initial conditions using 8 times higher mass resolution. The higher-resolution simulations of z5m10a and z5m11c have been first presented in \citet{Ma:2020a}.

The zoom-in regions are centered around halos chosen at desired mass and redshift from a set of dark matter (DM)-only cosmological boxes. The multi-scale cosmological zoom-in initial conditions are generated at $z=99$ using the {\sc music} code \citep{Hahn:2011} following the method from \citet{Onorbe:2014}. We ensure no contamination from low-resolution particles within $2\Rvir$ of the central halo, and less than 1\% contamination by mass within $3\Rvir$. 22 zoom-in regions are selected from a $(30\,\hinv\Mpc)^3$ box run to $z=5$ around halos in $\Mhalo\sim10^{9.5}$--$10^{12}\,\Msun$, 6 others are selected from a $(120\,\hinv\,\Mpc)^3$ box run to $z=7$, and the rest 6 from an independent box of the same size run to $z=9$. They are centered on halos from $\Mhalo\sim10^{11}$--$10^{12}\,\Msun$ at $z=7$ and $z=9$, respectively.

The initial mass for baryonic particles (gas and stars) is $\mb\sim100$, 900, or $7000\,\Msun$, and high-resolution DM particles $\mDM\sim650$--$4\times10^4\,\Msun$ in our simulations. Force softening for gas particles is adaptive, with a minimum Plummer-equivalent force softening length $\eg=0.14$--$0.42\,\pc$. Force softening lengths for star particles and high-resolution DM particles are fixed at $\es=5\eg=0.7$--2.1\,pc and $\eDM=10$--42\,pc, respectively. The softening lengths are in comoving units at $z>9$ and in physical units thereafter. In Table \ref{tbl:sims}, we provide the final redshift, mass resolution, force softening lengths, final halo mass, and stellar mass for all 34 zoom-in simulations analyzed in this paper. 

\begin{figure}
\centering
\includegraphics[width=\linewidth]{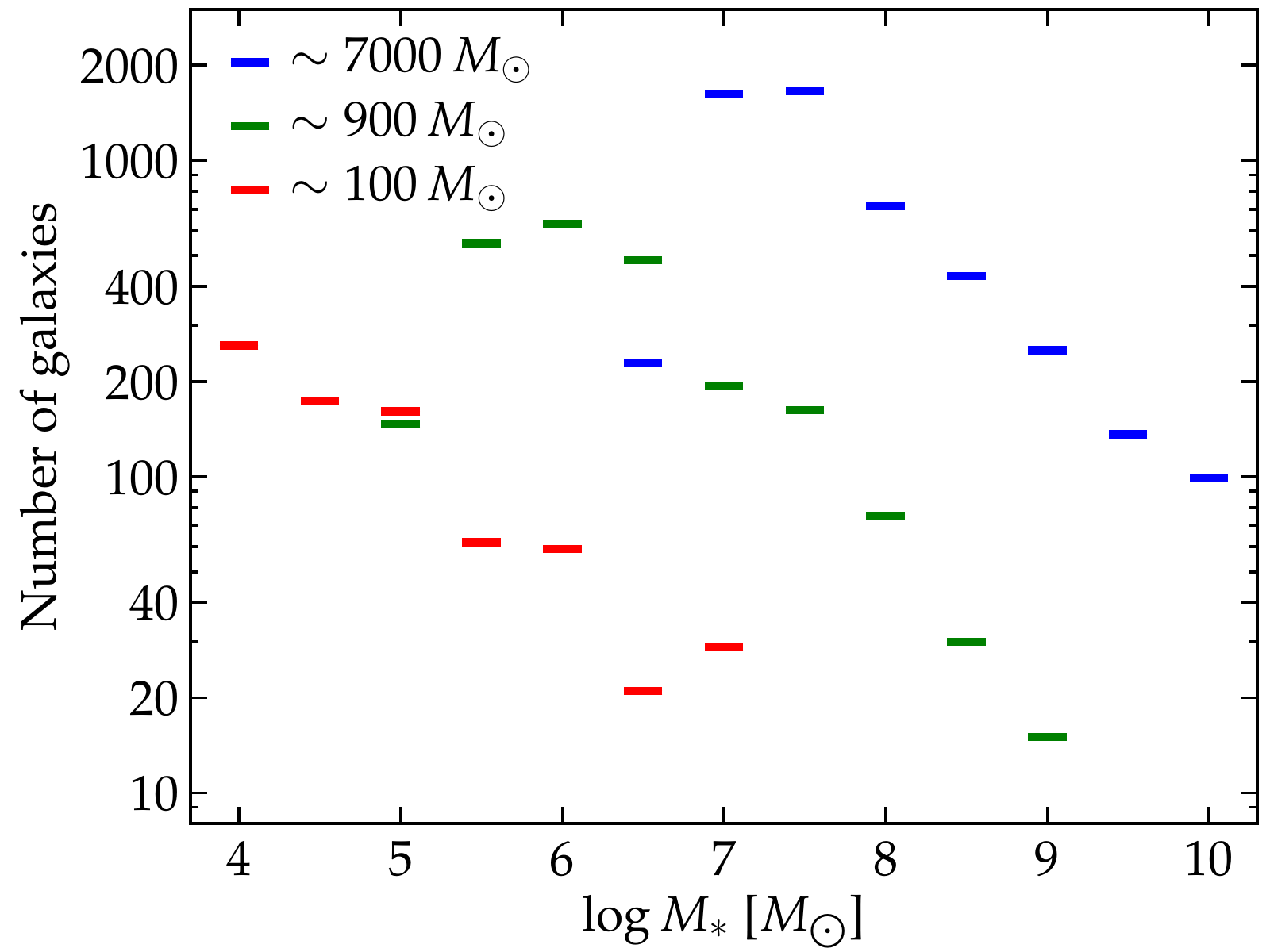}
\caption{Number of galaxies for every 0.5\,dex in stellar mass from $\Ms=10^{3.75}$--$10^{10.25}\,\Msun$ in our simulation sample. We show the sample size for the three resolution levels but at all redshift.} 
\label{fig:sampMs} 
\end{figure}

All simulation are run using an identical version of the code {\sc gizmo} \citep{Hopkins:2015} in the meshless finite-mass (MFM) mode with the FIRE-2 models of the multi-phase ISM, star formation, and stellar feedback \citep{Hopkins:2018b}, which we briefly summarize below. Gas follows an ionized+atomic+molecular cooling curve in 10--$10^{10}$\,K, including metallicity-dependent fine-structure and molecular cooling at low temperatures and high-temperature metal-line cooling for 11 separately tracked species (H, He, C, N, O, Ne, Mg, Si, S, Ca, and Fe). At each timestep, the ionization states and cooling rates for H and He are computed following \citet{Katz:1996} and cooling rates from heavier elements are calculated from a compilation of {\sc cloudy} runs \citep{Ferland:2013}, where we apply a uniform, redshift-dependent ionizing background from \citet{Faucher-Giguere:2009}\footnote{The ionizing background makes reionization complete at $z\sim9$ under the optically thin assumption.} and an approximate model for H\,{\sc ii} regions generated by local sources. Gas self-shielding is accounted for with a local Jeans-length approximation. 

Star formation is allowed only in dense, molecular and locally self-gravitating regions with hydrogen number density above $\nc=1000\,\cm^{-3}$ at 100 per\,cent efficiency per local free-fall time \citep[$\dot{\rho}_{\ast}=\epsilon\rho/t_{\rm ff}$, where $\epsilon=1$ by default; see][]{Hopkins:2013b}. Every star particle is regarded as a stellar population with known mass, age, and metallicity assuming a \citet{Kroupa:2002} IMF from 0.1--$100\,\Msun$. The simulations account for the following feedback mechanisms: (i) local and long-range radiation pressure,\footnote{This refers to single scattering of UV-optical radiation by dust and multiple scattering of reprocessed infrared photons in the optically-thick regime. Escaped flux is attenuated by the line-of-sight optical depth computed with a tree method (see appendix E in \citealt{Hopkins:2018b} for details).} (ii) photoionization and photoelectric heating, and (iii) energy, momentum, mass, and metal injections from SNe and stellar winds. The luminosities, mass loss rates, and Type-II SN rates for each star particle are obtained from {\sc starburst99} \citep{Leitherer:1999}, and Type-Ia SN rates from \citet{Mannucci:2006}. These simulations include a sub-resolution turbulent metal diffusion model described in \citet{Su:2017} and \linebreak \citet{Escala:2018}. We do not consider primordial chemistry nor Pop\,{\sc iii} star formation, but adopt an initial metallicity floor of $Z=10^{-4}\,\Zsun$. We emphasize that our photoionization feedback only includes a very approximate model for {\HII} regions: from each newly formed star particle, we move radially outwards and ionize neutral gas particles one by one until the the ionizing photon budget of the \linebreak star is exhausted; ionized gas is set to $10^4$\,K and photoelectric heating is included following the rates from \citet{Wolfire:2003}. \citet{Hopkins:2020} show that this approximate model does not make a big difference on galaxy-scale dynamics compared to more accurate radiation-hydrodynamic method even in halos down to $\Mhalo\sim$ \linebreak $10^9\,\Msun$ by $z=0$. In this work, we use post-processing MCRT calculations (Section \ref{sec:mcrt}) to obtain more accurate solutions of the ionization states, but we caution that the dynamical effect of photoionization feedback may not be correctly captured.

We use the Amiga's halo finder \citep[{\sc ahf};][]{Knollmann:2009} to identify halos and galaxies in the snapshots, applying the redshift-dependent virial parameter from \cite{Bryan:1998}. There are more than one halo in each zoom-in region. In this work, we restrict our analysis to halos that have zero contamination from low-resolution particles within $\Rvir$ and contain more than $10^5$ DM particles to ensure sufficient resolution for calculating $\fesc$. We also do not consider subhalos independently in this work. We utilize all snapshots we saved in our analysis and treat the same halo across different snapshots as different `galaxies'. The typical time separation is $\sim16$\,Myr between two consecutive snapshots. This is to account for short-time-scale variabilities of galaxy properties due to bursty star formation \citep[e.g.][]{Ma:2018a} and to maximize the statistical power of our simulation sample. Fig. \ref{fig:sample} shows the number of halo snapshots in our simulated sample in bins of 0.3\,dex in halo mass from $\Mhalo=10^{7.8}$--$10^{12}\,\Msun$ for three redshift bins ($5\leq z<7$, $7\leq z<9$, and $z\geq9$) at any resolution (top) and for the three mass \linebreak resolution levels at all redshifts (bottom). Fig. \ref{fig:sampMs} shows the number of snapshots in bins of 0.5\,dex in stellar mass from $\Ms=10^{3.75}$--$10^{10.25}\,\Msun$ for the three resolution levels at all redshifts. There are 8500 `galaxies' in total from $z=5$--12 in our simulated sample. 

\subsection{Monte Carlo radiative transfer of ionizing radiation}
\label{sec:mcrt}
We post-process every galaxy snapshot using a three-dimensional MCRT code to solve ionizing photon transport and ionization balance, from which we obtain $\fesc$ as a product. For every simulated galaxy, we map all gas particles within $\Rvir$ onto an octree grid: we first deposit all particles in a cubic root cell of side length equal to $2\Rvir$ and adaptively divide a parent cell into eight child cells until no leaf cell contains more than 2 gas particles. The minimum cell size in the densest region is usually less than 1\,pc even for the most massive galaxies in our sample. All physical quantities of a cell are evaluated using 32 nearest gas particles smoothed by a cubic spline kernel. We include all star particles within $\Rvir$ in our calculations. The hydrogen ionizing photon production rate of each star particle is computed from its age and metallicity using the Binary Population and Spectral Synthesis (BPASS) model \citep[v2.2.1;][]{Eldridge:2017}. We consider both the single-star and binary models and will compare the results in Section \ref{sec:binary}. The BPASS single-star models are very close to the {\sc starburst99} models adopted in our simulations for stellar feedback. We only consider binary models in post- processing, but we do not expect them to have a large effect on the feedback \citep[e.g.][]{Hopkins:2018b}.

\begin{table}
\centering
\caption{Three choices of stellar population and dust models we adopt for our Monte Carlo radiative transfer calculations.}
\begin{threeparttable}
\begin{tabular}{ccc}
\hline
Model & Stellar population model$^a$ & Dust model$^b$ \\
\hline
I (default) & single-star & default \\
II & binary & default \\
III & single-star & no dust \\
\hline
\end{tabular}
\begin{tablenotes}
\item $^a$ Both single-star and binary models are taken from the BPASS models (version 2.2.1).
\item $^b$Our default dust model adopts a constant dust-to-metal ratio of 0.4 in gas below $10^6$\,K (no dust at higher temperatures) with opacity $3\times10^5\,\cm^2\,\g^{-1}$ and albedo 0.277 at the Lyman limit.
\end{tablenotes}
\end{threeparttable}
\label{tbl:model}
\end{table}

The structure of our MCRT code is similar to those described in previous works \citep[e.g.][]{Fumagalli:2011,Ma:2015,Smith:2019}. A total number of 1--$2.4\times10^8$ photon packets are emitted isotropically from the location of star particles, sampled by their ionizing photon emissivity. The same number of photon packets are sent from domain boundary inwards to create an isotropic, uniform ionizing field with intensity given by \citet{Faucher-Giguere:2009}. Each photon packet is propagated until it escapes the domain, or is absorbed. The number of photon packets we use is sufficiently large such that it does not affect our results on $\fesc$. These photon packets are used to construct the ionization radiation field in the domain.

A photon packet may be absorbed by a neutral hydrogen atom with photoionization cross section from \citet{Verner:1996}. It may also be absorbed or scattered by dust grains. In our default calculations, we assume (a) 40\% of the metals are locked in dust grains in gas below $10^6$\,K while no dust in gas at higher temperature \citep[e.g.][]{Dwek:1998} and (b) the dust follows the Small Magellanic Cloud (SMC) grain-size distribution from \citet{Weingartner:2001}, which gives a dust opacity $3\times10^5\,\cm^2\,\g^{-1}$ and an albedo 0.277 at the Lyman limit.\footnote{It is worth pointing out that the effective dust cross section per hydrogen atom is $\sigma_{\rm d}\sim 4\times10^{-21}\,(Z/0.02)\,\cm^{-2}$ at the Lyman limit, where $Z$ is the gas-phase metallicity. This is much smaller than the neutral hydrogen cross section $\sigma_{\rm H\,I}\sim6.3\times10^{-18}\,\cm^{-3}$, which means dust is negligible in neutral gas but only important in mostly ionized gas  with column densities $N_{\rm H\,I}\gtrsim$ $2.5\times10^{20}\,(Z/0.02)^{-1}\,\cm^2$.} When the transport of all photon packets is done, we solve the ionization state for each cell assuming ionization equilibrium, where we adopt the temperature-dependent collisional ionization rates in \citet{Jefferies:1968} and recombination rates in \citet{Verner:1996a}. We take the gas temperatures in the simulations to compute these rates, since the simulations also take into account other heating sources (e.g. shocks) besides photo-heating. We iteratively transport photon packets and update the ionization states to reach convergence. We find 10 iterations sufficient for our purpose. In this paper, we consider three combinations of stellar population and dust models in our MCRT calculations as listed in Table \ref{tbl:model}. We focus on results from Model I, where we use single-star model and our default dust models, unless stated otherwise. We compare the results between these models in Section \ref{sec:fave}. 

In previous works, we have shown that our default dust model produces reasonable attenuation in the UV, bright-end UVLFs, and the relation between infrared excess and UV slope (i.e. the IRX--$\beta$ relation; see \citealt{Ma:2018a}, section 6 and \citealt{Ma:2019}). Nonetheless, this treatment is still simplistic and a detailed investigation on dust grain physics is beyond the scope of this paper. We consider a model where dust is ignored completely (Model III in Table \ref{tbl:model}) and compare the results with those in our default model to bracket the effect of dust on $\fesc$ (Section \ref{sec:dust}).

\section{The escape fraction of ionizing photons}
\label{sec:fesc}
In this section, we present the ionizing photon escape fraction, $\fesc$, for our simulated sample, its correlation with galaxy mass and redshift, and the effects of stellar population and dust on these results. We reiterate the fact that the $\fesc$ of a galaxy is defined as the absolute fraction of ionizing photons that escape the virial radius of the \linebreak halo. We investigate the key physics that governs the escape of ionizing photons in Section \ref{sec:physics}.

\begin{figure}
\centering
\includegraphics[width=\linewidth]{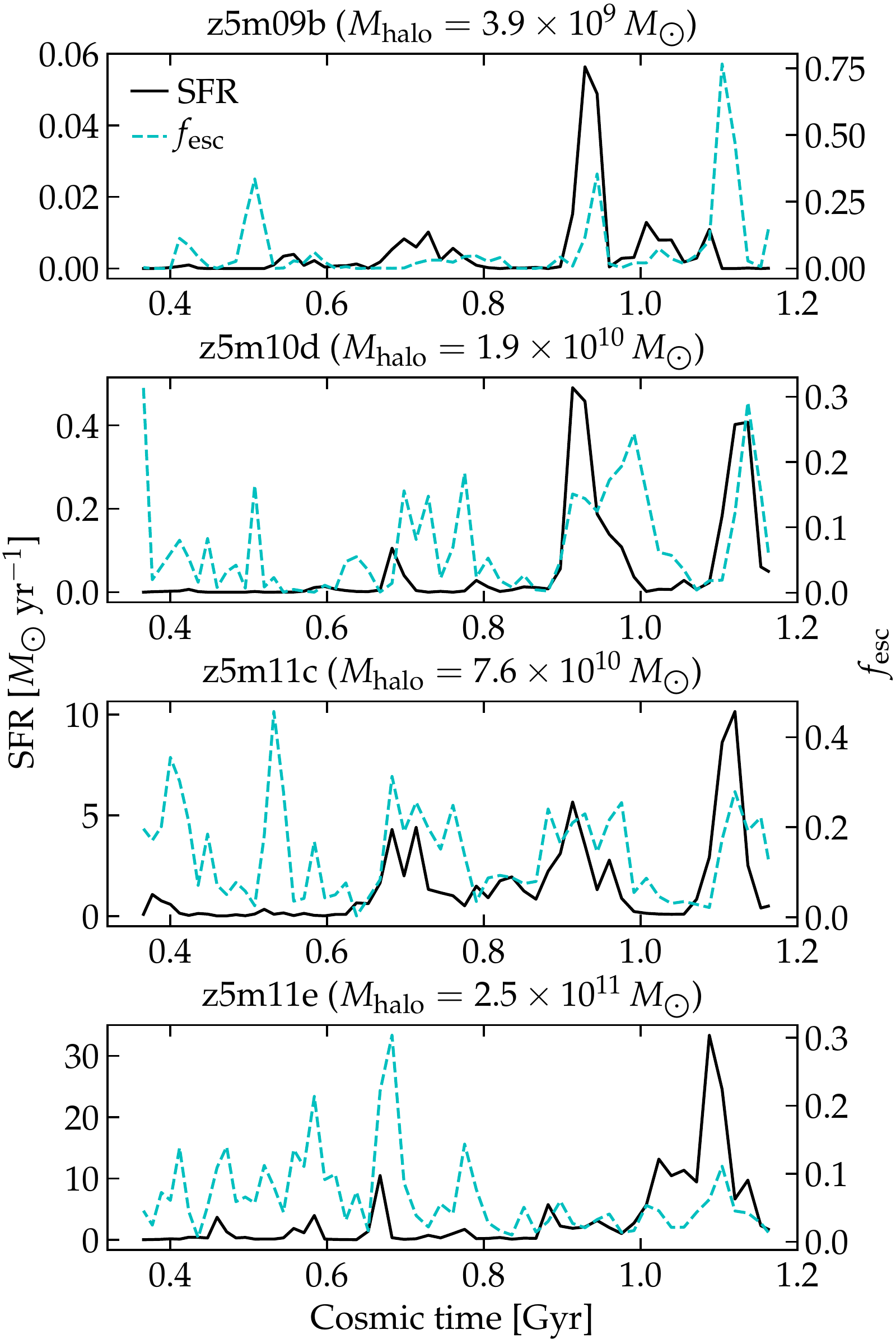}
\caption{The star formation rate (SFR; black solid lines) and instantaneous escape fraction ($\fesc$; cyan dashed lines) for four galaxies in our simulations, using single-star and our default dust models. Both the SFR and $\fesc$ show large variabilities on short time-scales. There is usually a time lag between the rising of $\fesc$ and the rising of SFR at the beginning of a starburst, as it takes some time for feedback to clear the sightlines where ionizing photons can escape. Note that a galaxy may have high $\fesc$ but low SFR at certain epochs, meaning that it is not leaking a large number of ionizing photons.}
\label{fig:sfr} 
\end{figure}

\begin{figure}
\centering
\includegraphics[width=\linewidth]{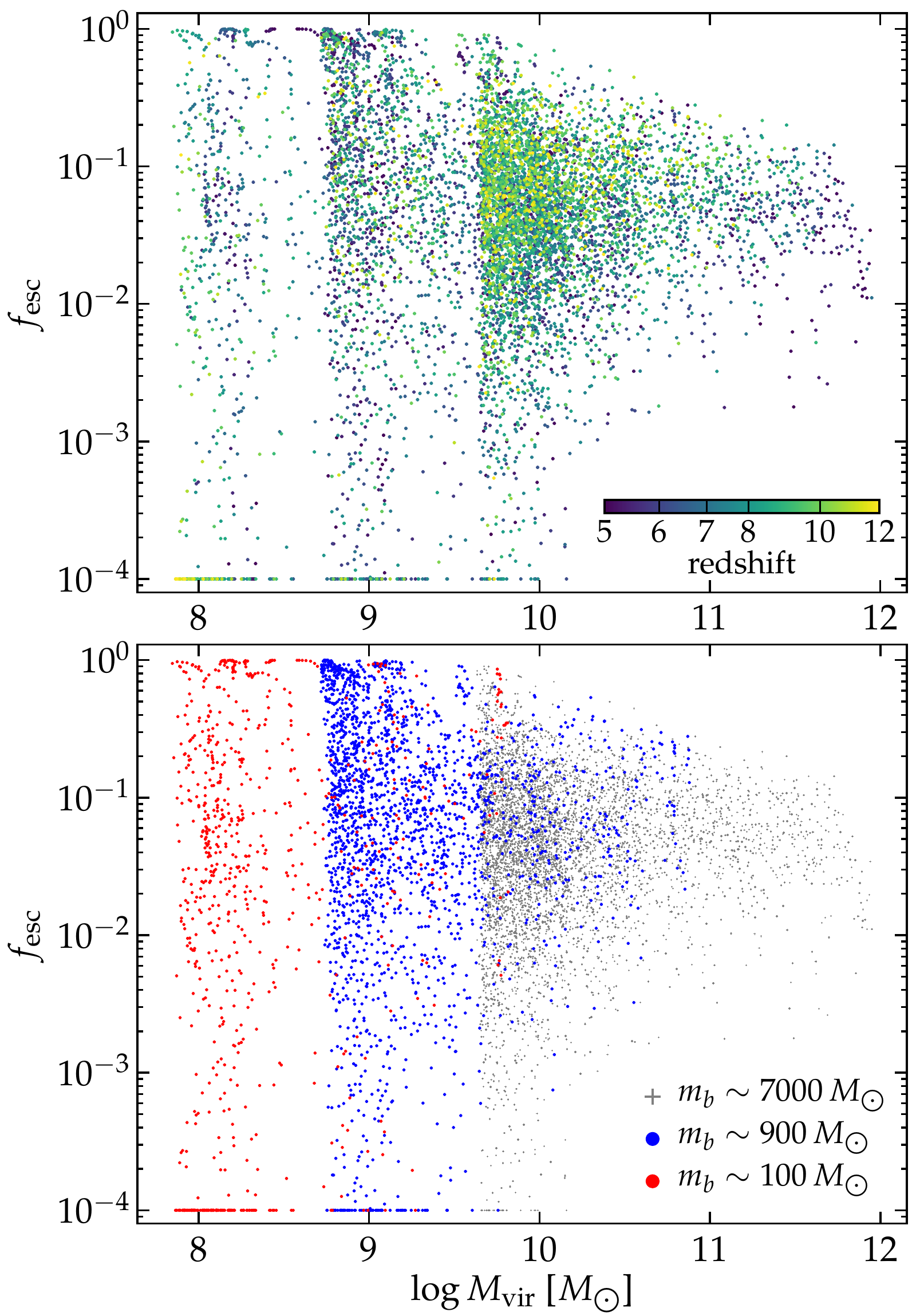}
\caption{The $\fesc$--$\Mvir$ relation (using single-star and default dust model). Each point shows a halo snapshot in our simulated sample. In the top panel, \newline the points are color-coded by redshift. In the bottom panel, the colors represent the mass resolution of the simulation. The instantaneous $\fesc$ has a large scatter (2--4\,dex) at a given $\Mvir$. Again, a high $\fesc$ does not necessarily mean the galaxy is leaking a large number of ionizing photons, as the SFR may be low at these epochs.}
\label{fig:fesc} 
\end{figure}

\subsection{The instantaneous escape fraction}
\label{sec:fins}
Fig. \ref{fig:sfr} shows the star formation rate (SFR; black solid lines) and $\fesc$ (cyan dashed lines) as a function of cosmic time for four examples in our sample: the central halos in z5m09b, z5m10d, z5m11c, and z5m11e, with halo mass spanning in $\Mvir\sim10^{9.6}$--$10^{11.5}\,\Msun$ by $z=5$. For each simulation, we show the 57 epochs from $z=12$ to 5 at which we saved a snapshot. Both $\fesc$ and the SFR show large time variabilities, with $\fesc$ changing from nearly zero to order unity (e.g. $\gtrsim20\%$) on short time-scales. There is usually a time delay between the rising of $\fesc$ and the rising of SFR when a starburst begins (e.g. z5m09b at $t\sim0.9$\,Gyr, z5m10d, and z5m11c at $t\sim1.1$\,Gyr). This has been reported in our previous works \citep[e.g.][]{Ma:2015,Smith:2019} and is because it takes a few Myrs for stellar feedback to clear some sightlines before a considerable fraction of the ionizing photons are allowed to escape. In Section \ref{sec:geo}, we further show how feedback determines the escape of ionizing photons.

Fig. \ref{fig:fesc} shows the relation between $\fesc$ and halo mass $\Mvir$ for our simulated sample. Each point represents a halo snapshot. In the top panel, the color shows the redshift of the snapshot, while in the bottom panel, we separate the points by the mass resolution we use for each simulation as $m_b\sim7000\,\Msun$ (grey), $900\,\Msun$ (blue), and $100\,\Msun$ (red). If a galaxy has $\fesc<10^{-4}$, we plot it at $10^{-4}$. At a given $\Mvir$, $\fesc$ shows a large scatter, spanning from less than $10^{-4}$ to unity. We caution that a galaxy with instantaneous $\fesc\sim1$ does not mean it is leaking a large number of ionizing photons, as seen from Fig. \ref{fig:sfr} that there are some epochs at which $\fesc$ is high but the SFR is very low. This happens more often in low-mass halos (or at early times), likely because feedback has blown out most of the gas at such epochs. Galaxies at different redshift, or run with different resolution, overlap largely on the $\fesc$--$\Mvir$ plane. At $\log\Mvir\gtrsim10$, the smaller scatter is likely due to weaker bursty star formation at higher masses \citep[e.g.][]{Hopkins:2014,Kimm:2014}. The scatter in this relation is too large to reveal any significant correlation of $\fesc$ with $\Mvir$ or redshift, or the robustness of our results to mass resolution. We address these questions by studying the sample averaged $\fave$ in Section \ref{sec:fave}.

\begin{figure*}
\centering
\includegraphics[width=\linewidth]{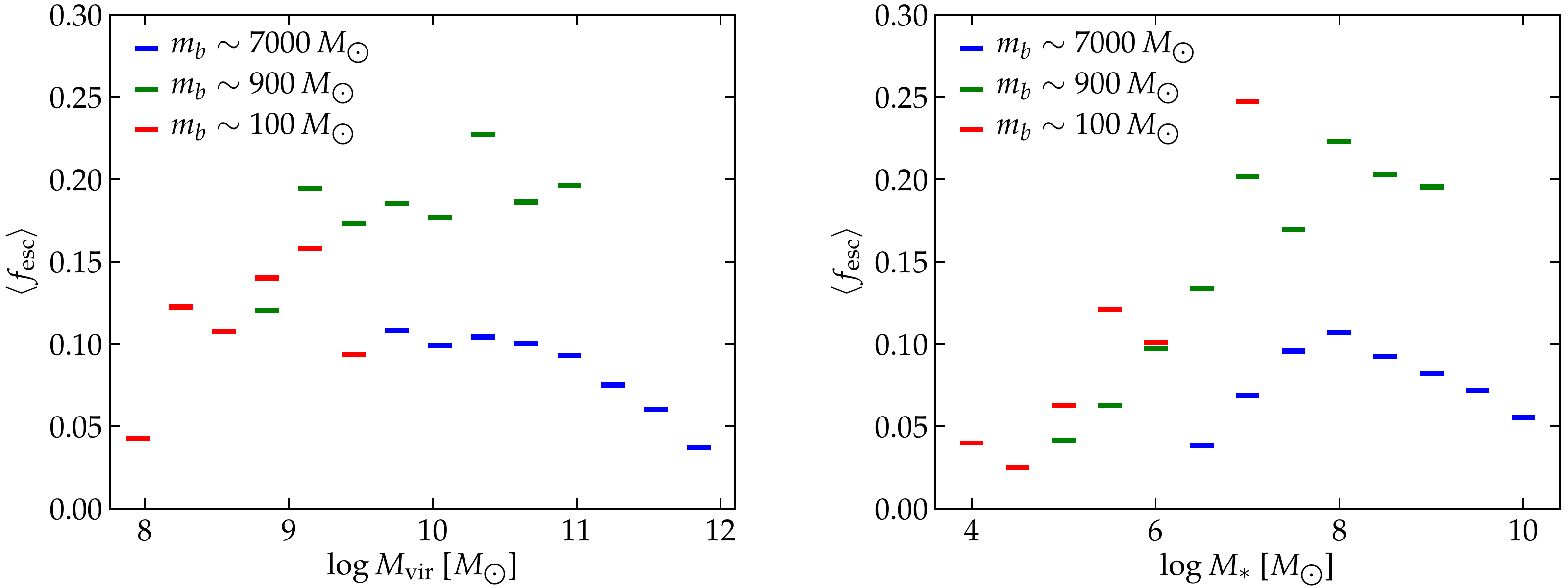}
\caption{The sample average $\fave$ as a function of $\Mvir$ (left) and $\Ms$ (right) (using single-star and default dust model). $\fave$ is defined as the $Q_{\rm ion}$-weighted instantaneous $\fesc$ over all snapshots at all redshifts in a given halo/stellar mass bin. The blue, green, and red bars represent simulations at $m_b\sim7000$, 900, and $100\,\Msun$ resolution. Our results do not fully converge with resolution. Simulations at $m_b\sim7000\,\Msun$ resolution tend to produce lower $\fave$ than simulations at $m_b\sim900\,\Msun$ or better resolution. $\fave$ increases with $\Mvir$ in $\log\Mvir\sim8$--9.5, becomes nearly independent of $\Mvir$ in $\log\Mvir\sim9.5$--11, and decreases with $\Mvir$ above $\log\Mvir\sim11$. $\fave$ increases with $\Ms$ up to $\log\Ms\sim8$ and decreases with $\Ms$ at higher masses. The decrease of $\fesc$ at the high- and low-mass end is due to increasingly strong dust attenuation (cf. Fig. \ref{fig:fMcomp} and Section \ref{sec:dust}) and inefficient star formation and feedback (cf. Figs. \ref{fig:fage}--\ref{fig:halocol} and Section \ref{sec:col}), respectively. It is worth noting that the compromise between these two effects results in the apparent independence of $\fave$ on $\Mvir$ in $\log\Mvir\sim9.5$--11.}
\label{fig:fescM}
\end{figure*}

\subsection{The average escape fraction}
\label{sec:fave}
\subsubsection{Correlations with galaxy mass}
\label{sec:fM}
We divide our sample into 14 equal-width bins in logarithmic halo mass from $\log\Mvir=7.8$--12, with a bin width of 0.3\,dex. For each bin, we compute the average escape fraction over all halo snapshots in that bin, $\fave=\sum_i Q_{{\rm esc},\,i} / \sum_i Q_{{\rm ion},\,i} = \sum_i f_{{\rm esc},\,i} \, Q_{{\rm ion},\,i} / \sum_i Q_{{\rm ion},\,i}$, where $Q_{{\rm ion},\,i}$ and $Q_{{\rm esc},\,i}$ are the number of ionizing photons emitted and escaped per unit time, and $f_{{\rm esc},\,i}$ is the escape fraction of the $i^{\rm th}$ galaxy in the bin. In the left panel of Fig. \ref{fig:fescM}, we show the correlation between $\fave$ and halo mass $\Mvir$. The color separates simulations run with different resolution (blue: $m_b\sim7000\,\Msun$, green: $900\,\Msun$, and red: $100\,\Msun$). Here we average over galaxies at all redshifts in a given mass bin, but we study redshift dependence later in Section \ref{sec:redshift}. Note that certain bins do not have a large number of galaxies ($\lesssim100$, see Fig. \ref{fig:sample}), which may introduce noise to our results.

\begin{figure*}
\centering
\includegraphics[width=\linewidth]{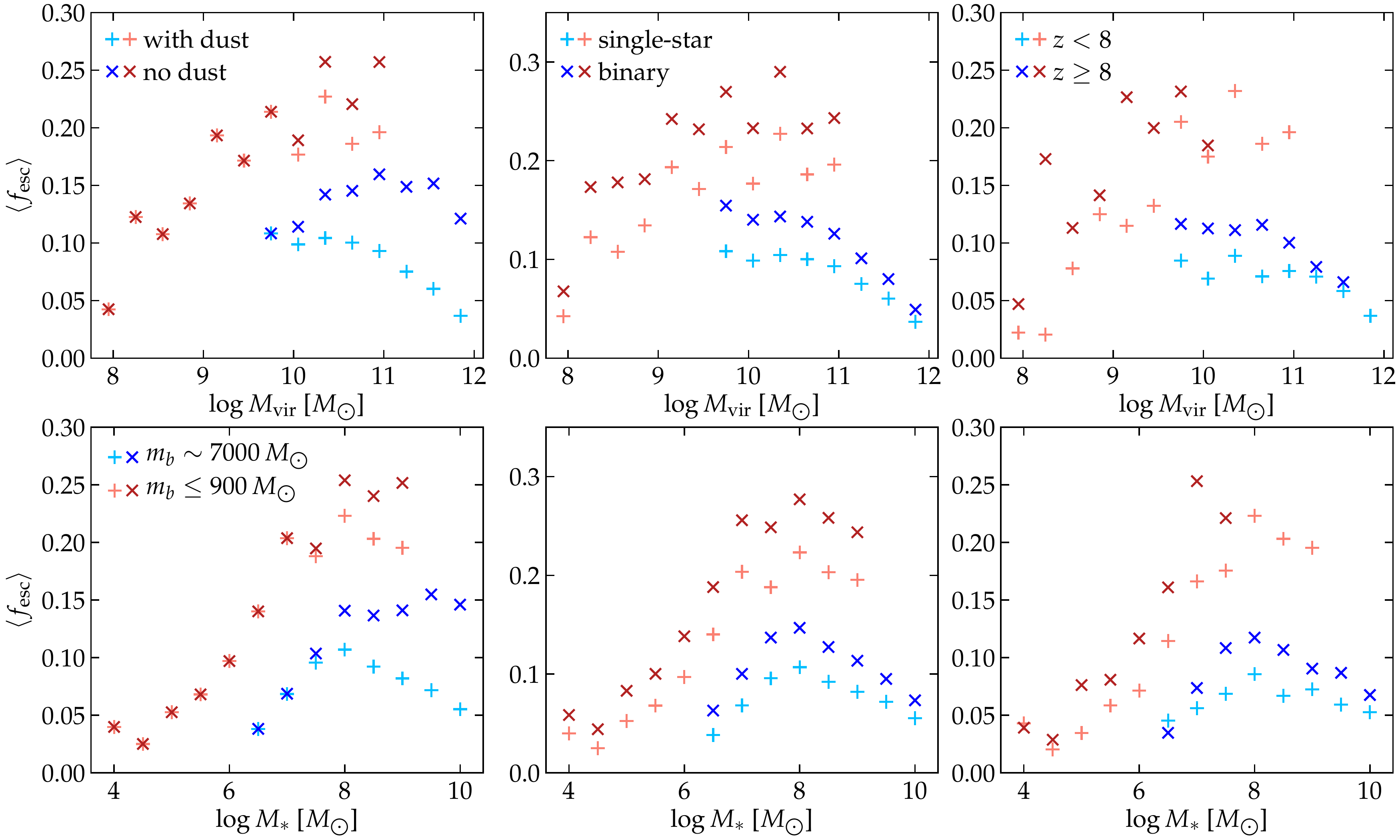}
\caption{The $\fave$--$\Mvir$ (top) and $\fave$--$\Ms$ (bottom) relation, separated by simulations at $7000\,\Msun$ resolution (blue) and at $900\,\Msun$ resolution or better (red). {\em Left column}: $\fave$ with (light plus signs) and without (dark cross signs) dust attenuation. $\fave$ increases with halo (stellar) mass up to $\log\Mvir\sim11$ ($\log\Ms\sim8$) and becomes roughly constant at higher masses without dust attenuation, suggesting the decrease of $\fave$ at the high-mass end in Fig. \ref{fig:fescM} is due to dust attenuation (Section \ref{sec:dust}). {\em Middle column}: $\fave$ calculated using single-star (plus) and binary (cross) stellar population models. Binary models tend to boost $\fave$ by $\sim25$--35\% in most mass bins, as they predict more ionizing photons for a stellar population after 3\,Myr and the extra photons have a higher possibility to escape (cf. Fig. \ref{fig:fage} and Section \ref{sec:age}). Binary stars increase the number of ionizing photons {\em emitted} by $\sim20$--30\%, and thus the number of ionizing photons {\em escaped} by $\sim60$--80\% (Section \ref{sec:binary}). {\em Right column}: The dependence of $\fave$ on redshift. The sample is divided into two redshift bins: $z<8$ (plus) and $z\geq8$ (cross). Galaxies at higher redshifts tend to have systematically higher $\fave$ than galaxies at lower redshifts (Section \ref{sec:redshift}).}
\label{fig:fMcomp} 
\end{figure*}

Our results on $\fave$ do not yet fully converge with resolution. Simulations at $7000\,\Msun$ resolution tend to produce systematically lower $\fave$ than those at $900\,\Msun$ resolution or better, while we do not find significant differences between simulations at $900\,\Msun$ and $100\,\Msun$ resolution. Now we focus on the qualitative trend between $\fave$ and $\Mvir$. For intermediate halo mass (i.e. $\log\Mvir\sim9.5$--11), $\fave$ is nearly constant and does not depend strongly on halo mass. This trend is found both for simulations at $7000\,\Msun$ resolution and at $900\,\Msun$ or better resolution, although the absolute value of $\fave$ differs between the two subsamples. In this halo mass range, $\fave$ \linebreak is roughly 0.2 for simulations at $900\,\Msun$ resolution, while $\fave\sim0.1$ for those at $7000\,\Msun$ resolution. At the massive end (i.e. above $\log\Mvir\sim11$), $\fave$ decreases with halo mass and drops to 0.03 at $\log\Mvir\sim12$. This is due to increasingly important dust attenuation at higher masses (cf. Fig. \ref{fig:fMcomp} and Section \ref{sec:dust}). Below $\log\Mvir\sim9$, $\fave$ decreases with decreasing halo mass and drops under 0.05 at \linebreak $\log\Mvir\sim8$. We discuss in Section \ref{sec:col} that this is likely due to less efficient star formation and stellar feedback in low-mass galaxies in clearing the gas for ionizing photons to escape. We emphasize that the apparent independence of $\fave$ on $\Mvir$ from $\log\Mvir\sim9.5$--11 is likely a coincidence: we show below in Section \ref{sec:dust} that if there is no dust, $\fave$ will increase with $\Mvir$ up to $\Mvir\sim10^{11}\,\Msun$ (cf. Fig. \ref{fig:fMcomp}), and the effect of dust attenuation becomes increasingly strong at higher masses, leading to a nearly constant $\fave$ at intermediate halo mass.

We also bin our sample in every 0.5\,dex in stellar mass into 13 equal-width stellar mass bins from $\log\Ms\sim3.75$--10.25. We show in the right panel of Fig. \ref{fig:fescM} the correlation between $\fave$ and stellar mass, where we average $\fesc$ over galaxies at all redshifts in a given stellar mass bin and the color represents the resolution used for our simulations. Again, simulations run at $7000\,\Msun$ resolution produce systematically lower $\fave$ than those run at $900\,\Msun$ resolution or better, but the qualitative behavior of the $\fave$--$\Ms$ relation agrees well between the two subsamples. We find that that $\fave$ increases with stellar mass until $\log\Ms\sim8$, where $\fave$ starts to decrease at higher masses. We obtain $\fave\sim0.2$ (0.1) for $900\,\Msun$ ($7000\,\Msun$) resolution at $\log\Ms\sim8$ where $\fave$ peaks. Similar to what mentioned above, the low $\fave$ at the high- and low-mass end is due to heavy dust attenuation and inefficient feedback, respectively, which we explicitly show later in this paper.

We reiterate the facts that simulations at $7000\,\Msun$ resolution and those at $900\,\Msun$ or better resolution predict broadly consistent \linebreak trends between $\fave$ and $\Mvir$ ($\Ms$) where the mass overlaps. Also, simulations at $900\,\Msun$ and $100\,\Msun$ resolution predict broadly similar $\fave$. This suggests that the qualitative trend in the $\fave$--$\Mvir$ ($\Ms$) relation from Fig. \ref{fig:fescM} is likely robust and not an artifact due to the non-trivial selection criteria for our simulated sample.

\subsubsection{The effects of dust attenuation}
\label{sec:dust}
For every halo above $\Mvir=10^{10}\,\Msun$ in our simulations, we repeat the MCRT calculations without dust extinction and scattering (`no dust', i.e. Model III in Table \ref{tbl:model}). We expect dust to be subdominant in halos below $\Mvir=10^{10}\,\Msun$, as they are much less dust-enriched than more massive halos. In the left column of Fig. \ref{fig:fMcomp}, we compare the $\fave$--$\Mvir$ relation (top panel) and the $\fave$--$\Ms$ relation (bottom panel) with and without dust attenuation. Blue and red symbols represent simulations run at $7000\,\Msun$ resolution and at $900\,\Msun$ or better resolution, respectively. Thereafter, we combine simulations at $900\,\Msun$ and $100\,\Msun$ resolution, given that we do not find a significant difference between the two resolution and our sample does not contain a sufficiently large number of galaxies at $100\,\Msun$ resolution. The light plus and dark cross signs show the results from Model I (with our default dust model) and Model III (without dust attenuation), respectively. Combining the qualitative trend revealed by the two subsamples at $\sim7000\,\Msun$ and at $\lesssim900\,\Msun$ resolution, we find that without dust attenuation, $\fave$ increases with $\Mvir$ and $\Ms$ until $\Mvir\sim10^{11}\,\Msun$ and $\Ms\sim10^8\,\Msun$ to 0.25 (0.15) for simulations at $900\,\Msun$ ($7000\,\Msun$) resolution, and turns nearly constant at the more massive end. This suggests that the seemingly constant $\fave$ over $\log\Mvir\sim9.5$--11 and the decrease of $\fave$ at the high-mass end ($\log\Mvir\gtrsim11$, $\log\Ms\gtrsim8$) are due to increasingly heavy dust attenuation.

\subsubsection{The effects of binary stars}
\label{sec:binary}
We also repeat our MCRT calculations with the binary stellar population models instead of the single-star models from BPASS (i.e. Model II in Table \ref{tbl:model}). The binary models include mass transfer from the primary star to the secondary star and binary merger that make more high-mass stars at later times compared to single-star models. They also take into account quasi-homogeneous evolution for low-metallicity, fast rotating stars (due to mass transfer), whose surface temperatures are high \citep[see e.g.][]{Eldridge:2012,Eldridge:2017}. For a single-age stellar population, the ionizing photon emissivity is nearly the same between single-star and binary models in the first 3\,Myr, but binary models predict more ionizing photons than single-star models after 3\,Myr due to binary evolution. Binary models produce $\sim20$--35\% more ionizing photons for a single-age population of 0.1--$10^{-3}$ solar metallicity over its lifetime. Besides, feedback is expected to clear a large fraction of the sightlines after 3\,Myr, so the extra ionizing photons from binary stars are likely to escape more easily (see also Section \ref{sec:age}). These two effects suggest that binary stars may contribute a large number of ionizing photons for reionization \citep[e.g.][]{Ma:2016,Stanway:2016,Rosdahl:2018,Gotberg:2017,Gotberg:2020}.

In the middle column of Fig. \ref{fig:fMcomp}, we compare the results using single-star (light plus signs) and binary (dark cross signs) models. We restate that the binary models are only used in post-processing calculations, not on-the-fly in our simulations. Again, the top panel and the bottom panel show the $\fave$--$\Mvir$ and $\fave$--$\Ms$ relations, respectively, and the colors represent the resolution for our simulations. We confirm previous works that binary stars tend to produce systematically higher $\fave$, although the qualitative trend between $\fave$ and halo/stellar mass remains the same as single-star models. However, we find that binary stars only boost $\fave$ moderately by $\sim25$--35\%, and the average ionizing photon production rate $\langle Q_{\rm ion}\rangle$ by $\sim20$--30\%, so the number of ionizing photons {\em escaped} per unit \linebreak time $\langle Q_{\rm esc}\rangle$ is boosted by about $\sim60$--80\% compared to single-star models for most halo mass and stellar mass bins. Our results here suggest that binary stars have smaller effects than previously found \citep[cf. a factor of 3 or more, e.g.][]{Ma:2016,Rosdahl:2018}, probably because stars younger than 3\,Myr (when binary evolution \linebreak is subdominant) leak ionizing photons more efficiently in our current FIRE-2 simulations, thereby lowering the relative contribution by the extra ionizing photons from binary stars. We further discuss this in Section \ref{sec:age}.

\subsubsection{Dependence on redshift}
\label{sec:redshift}
So far, we only studied the average escape fraction over galaxies at all redshifts. Now we investigate the redshift dependence of $\fave$. We note that our sample has a very small size, so we only divide it into two redshift bins: $z<8$ and $z\geq8$. There will be too few galaxies in many halo (stellar) mass bins if we use more redshift bins, so we only focus on the qualitative trend of $\fave$ with redshift. In the right column of Fig. \ref{fig:fMcomp}, we present the $\fave$--$\Mvir$ (top) and $\fave$--$\Ms$ (bottom) relations, where the plus and cross symbols represent galaxies at $z<8$ and $z\geq8$, respectively. The colors represent simulations at different resolution. For nearly all mass bins, $z\geq8$ galaxies show systematically higher $\fave$ than their $z<8$ counterparts. We speculate that this is because SFR increases with redshift for a given halo (stellar) mass \citep[e.g.][]{Ma:2018a}, so feedback is more efficient in clearing the gas and allowing more ionizing photons to escape owing to the stronger star formation activities at higher redshifts. A decreasing $\fave$ toward lower redshift has been proposed \linebreak in some models of the reionization history \citep[e.g.][]{Kuhlen:2012,Faucher-Giguere:2020,Yung:2020}. Finally, we stress that the qualitative behaviors of the $\fave$--$\Mvir$ and $\fave$--$\Ms$ relations are roughly the same between the two redshift bins.

\section{Physics of ionizing photon escaping}
\label{sec:physics}
In Section \ref{sec:fesc}, we show that although the instantaneous $\fesc$ of individual galaxies ranges from $\lesssim10^{-4}$ to 1, sample averaged $\fave$ is moderate, with $\fave\sim0.2$ around $\Ms\sim10^8\,\Msun$ using single-star stellar population models (for simulations at $900\,\Msun$ resolution or better). We reiterate that not every galaxy with a high instantaneous $\fesc$ is a strong leaker of ionizing photons. Only those that have both high $\fesc$ and SFR (in the middle of a starburst) are likely the dominant contributor to reionization. Moreover, we find that the average $\fave$ decreases at $\Ms\gtrsim10^8\,\Msun$ due to dust attenuation, but $\fave$ also decreases with decreasing stellar mass at $\Ms\lesssim10^8\,\Msun$. Note that strong leakers also exist, but are not common, among low-mass \linebreak galaxies. In this section, we address two questions: (1) in what conditions can ionizing photons escape efficiently, and (2) why do low-mass galaxies have low escape fractions on average?

Our analysis below mainly uses simulations at $900\,\Msun$ or better resolution, but we have confirmed that the qualitative behaviors are similar for simulations at $7000\,\Msun$ resolution. We mainly focus \linebreak on results using single-star models, in which almost all the ionizing photons are emitted from stars {\bf younger than 10\,Myr} (cf. Section \ref{sec:age}), unless stated otherwise.

\begin{figure*}
\centering
\includegraphics[width=\linewidth]{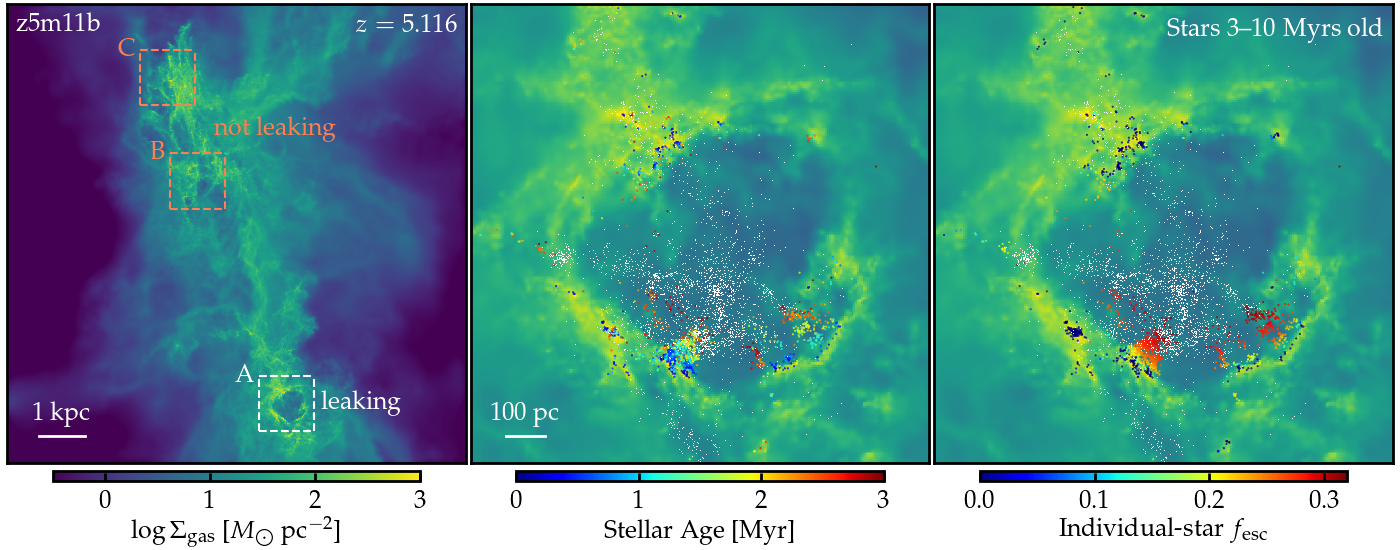}
\includegraphics[width=\linewidth]{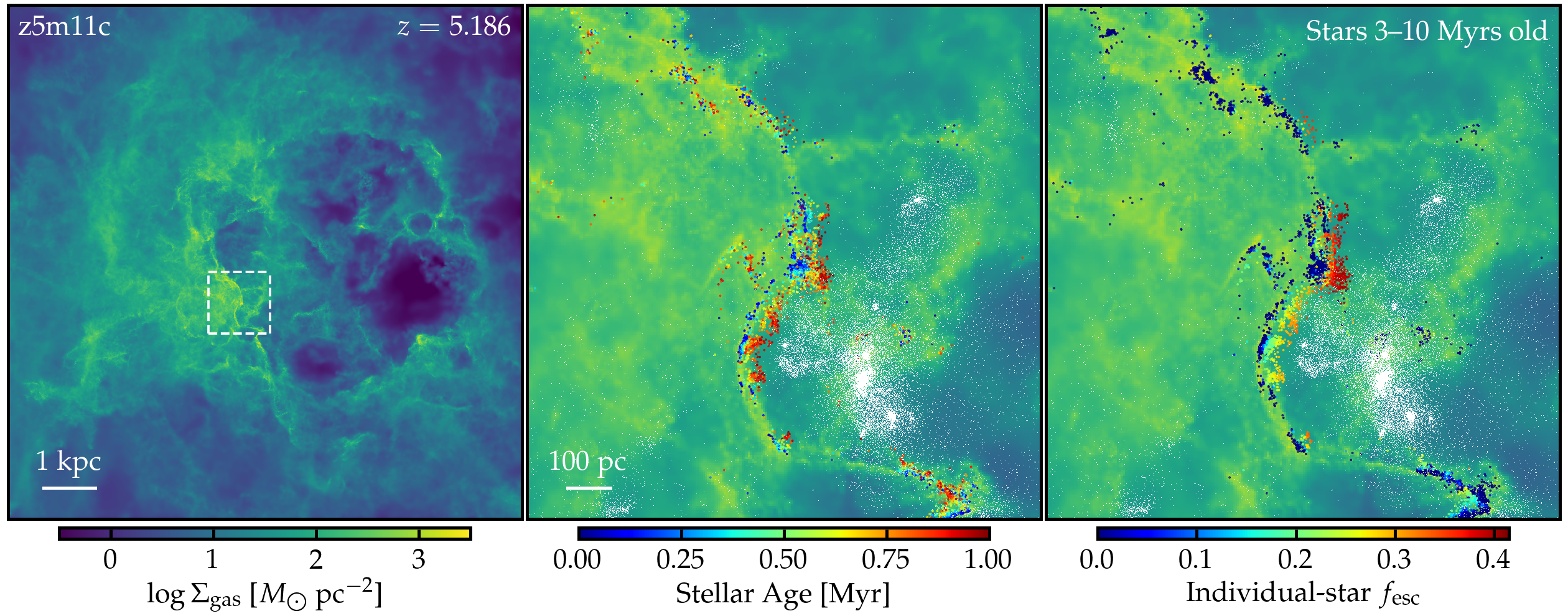}
\caption{Examples of galaxies with strong ionizing photon leakage. {\em Top row}: The central galaxy in simulation z5m11b at $z=5.116$. The galaxy is at the early stage of a starburst. At this epoch, it has a halo (stellar) mass of $\Mvir=3.7\times10^{10}\,\Msun$ ($\Ms=1.5\times10^8\,\Msun$) and instantaneous $\fesc\sim0.2$. {\em Left}: Gas surface density of a $10\,\kpc\times10\,\kpc$ projection. Most stars in the past 10\,Myr are formed in region A (marked by the white dashed square), where the majority of the {\em escaped} ionizing photons come from. Regions B and C have formed an order of magnitude fewer stars than region A, but almost no ionizing photons escape from both regions. {\em Middle}: Zoom-in image on region A (1.2\,kpc on each side). The white points show stars 3--10\,Myr old, while the color points show stars younger than 3\,Myr, color-coded by their ages. {\em Right}: The same image as in the middle panel, except the young stars are color-coded by their {\em single-star} escape fractions. Region A contains a kpc-scale superbubble presumably created by stars 3--10\,Myr old. A dense shell around the bubble is forming new stars while accelerated by feedback, leaving an age gradient at the bubble edge. Stars 2--3\,Myr old are already in the low-density bubble. The large number of young stars in region A can fully ionize the low-column-density sightlines around the bubble, allowing a large fraction of ionizing photons to escape. In contrast, regions B and C do not contain a feedback-driven superbubble nor a large number of young stars to fully ionize the surrounding gas. {\em Bottom row}: The central galaxy of simulation z5m11c at $z=5.186$. The galaxy is in the middle of a starburst. It has a halo (stellar) mass of $\Mvir=7.4\times10^{10}\,\Msun$ ($\Ms=7.8\times10^8\,\Msun$) and \newline instantaneous $\fesc\sim0.26$. The left panel shows the gas surface density of a $10\,\kpc\times10\,\kpc$ projection. The white dashed square marks an active star-forming region where most of the escaped ionizing photons come from. The middle and right panels show the zoom-in image on this region. The white points show stars 3--10\,Myr old and the color points show stars younger than 1\,Myr, color-coded by stellar age (middle) and single-star escape fraction (right). This region contains a dense shell around a superbubble of several kpc in size. The shell is forming stars while accelerated presumably by feedback from stars 3--10\,Myr nearby, so even stars 0.5--1\,Myr old are already inside the shell, leaking 30--40\% of their ionizing photons. We find such configuration (i.e. superbubble surrounded by a dense, accelerated shell) very common in strong ionizing-photon-leaking galaxies in our simulations, regardless of stellar mass and resolution. }
\label{fig:leak}
\end{figure*}

\subsection{Geometry of ionizing-photon-leaking regions}
\label{sec:geo}
In this section, we present some example galaxies with strong ionizing photon leakage from our simulations to establish an intuitive picture for the escaping of ionizing photons. This is also useful for understanding the LyC-leaking galaxies discovered at intermediate and low redshifts (see Section \ref{sec:intro} and references therein).

The top row of Fig. \ref{fig:leak} shows the central galaxy of simulation z5m11b at $z=5.116$. The halo mass and stellar mass at this epoch are $\Mvir=3.7\times10^{10}\,\Msun$ and $\Ms=1.5\times10^8\,\Msun$, respectively. It is at the early stage of starburst that begins 25\,Myr ago. The galaxy has an instantaneous $\fesc\sim0.2$ and rapidly rising SFR at this time. The left panel presents the gas surface density of a 10\,kpc$\times$10\,kpc region centered at the halo center. A large gas reservoir has built up in the ISM that triggered the starburst ($M_{\rm gas}\sim7.5\times10^8\,\Msun$ in the inner 5\,kpc). Almost all star formation in the past 10\,Myr happens in the three regions marked by the dashed squares (1.2\,kpc on each \linebreak side). Most stars were formed in region A (white). It is also where nearly all the {\em escaped} ionizing photons come from. Regions B and C (red) formed an order of magnitude fewer stars, and few ionizing photons from these stars can escape. 

Region A contains a kpc-scale superbubble surrounded by an incomplete dense shell. In the top-middle panel of Fig. \ref{fig:leak}, we zoom into the 1.2\,kpc$\times$1.2\,kpc region marked by box A. The white points show stars formed 3--10\,Myr ago, while the color points show stars younger than 3\,Myr, color-coded by their ages. The top-right panel shows the same image, except that the young stars are color-coded by the escape fraction of {\em individual stars}.\footnote{Thanks to the nature of the Monte Carlo method, we are able to track the source from which a photon packet is emitted in our MCRT calculations, so we know how many photon packets are emitted from each star particle and how many of them eventually escape. This should not be confused with the galaxy escape fractions (i.e. averaged over all stars in the galaxy).} The superbubble is presumably created by clustered SNe from stars 3--10\,Myr old.\footnote{The total mass of stars 3--10\,Myr old in the superbubble is $2.2\times10^6\,\Msun$.} These stars are sitting in the low-density bubble at this time. In the meanwhile, new stars form in the compressed, dense shell as the bubble expands in the ISM. More importantly, the shell can be accelerated while forming stars. This is the reason why there is an age gradient in stars younger than 3\,Myr at the bottom half of the shell \citep[see also][]{Yu:2020}. As a consequence, stars that are only 2--3\,Myr old already locate inside the low-density bubble. Moreover, the bubble is not completely covered by dense gas, with a large fraction of the sightlines cleared by feedback along which the gas column density is low (e.g. the direction pointing out of the image). The bubble has a large number of young stars. We will show in Section \ref{sec:feedback} that the low-column-density sightlines can be fully ionized by these young stars, allowing ionizing photons to escape effectively through these optically-thin channels. Stars 3--10\,Myr old in the bubble, and stars 2--3\,Myr old at the inner side of the bubble (which have $\fesc\sim0.3$), contribute the majority of the escaped ionizing photons.

To summarize, region A is leaking ionizing photons along the optically-thin sightlines around the superbubble. The escaped photons come from stars 3--10\,Myr old in the bubble and younger stars at the inner edge of the shell. The low-column-density channels are pre-cleared by feedback and then fully ionized by the large amount of young stars collectively in the bubble (cf. Fig. \ref{fig:col} and Section \ref{sec:feedback}). In contrast, regions B and C do not contain a superbubble. We find that most of the young stars in these regions are still buried in their birth clouds. Even stars 3--10\,Myr old are surrounded by optically-thick neutral gas in the ISM. This suggests that feedback in regions B and C has not been sufficiently strong to clear some channels that can be fully ionized to allow ionizing photons to escape.

The bottom row of Fig. \ref{fig:leak} shows another example, the central galaxy in simulation z5m11c at $z=5.186$, when the galaxy is at the peak of a starburst. The halo (stellar) mass is $\Mvir=7.4\times10^{10}\,\Msun$ ($\Ms=8\times10^8\,\Msun$) and instantaneous $\fesc\sim0.26$ at this time. The left panel shows the gas surface density in a 10\,kpc$\times$10\,kpc region around the halo center. The middle panel zooms into the $(1.2\,\kpc)^2$ region marked by the white dashed box in the left. This region is at the edge of a superbubble of a few kpc in size and contains a dense shell compressed by the bubble. This is the most active star-forming region in the past 10\,Myr, where the majority of the escaped ionizing photons come from. The white points show stars 3--10\,Myr old, which locate inside the low-density bubble. The color points show stars younger than 1\,Myr (rather than 3\,Myr in the top row), color-coded by their age in the middle panel and by $\fesc$ in the right panel. The shell is star-forming while accelerated presumably by the stars 3--10\,Myr nearby, leading to the age gradient at the edge of the bubble. The low-column-density sightlines can be fully ionized by the young stars in this region, allowing ionizing photons to escape from {\linebreak} these optically-thin paths. Even stars only 0.5--1\,Myr old are inside the inner edge of the shell (in the low-density bubble), making 30--40\% of their ionizing photons escape.

The two examples shown above share some similar features in regions that leak ionizing photons effectively. It must be an actively star-forming region that contains or is part of a kpc-scale superbubble. The bubble is surrounded by a compressed, dense shell where new stars form. The shell is usually accelerated, leaving an age gradient in the newly formed stars at the edge of the bubble. The bubble should not be fully confined, with low-column-density channels pre-cleared by feedback and fully ionized by the young stars in this region, from which ionizing photons can escape efficiently. Most of the escaped ionizing photons come from stars younger than 3\,Myr in the inner edge of the shell and stars 3--10\,Myr old inside the low-density bubble. We find such configuration very common in strong ionizing photon leakers (galaxies that have high $\fesc$ and SFR at the same time) in our simulations, for galaxies of all masses and run at any mass resolution.

\subsection{The important role of feedback}
\label{sec:feedback}
In the previous section, we use some examples to illustrate the typical geometry of strong ionizing-photon-leaking regions in galaxies with both high $\fesc$ and SFR. In this section, we investigate the key physics that governs the escape of ionizing photons.

\begin{figure}
\centering
\includegraphics[width=\linewidth]{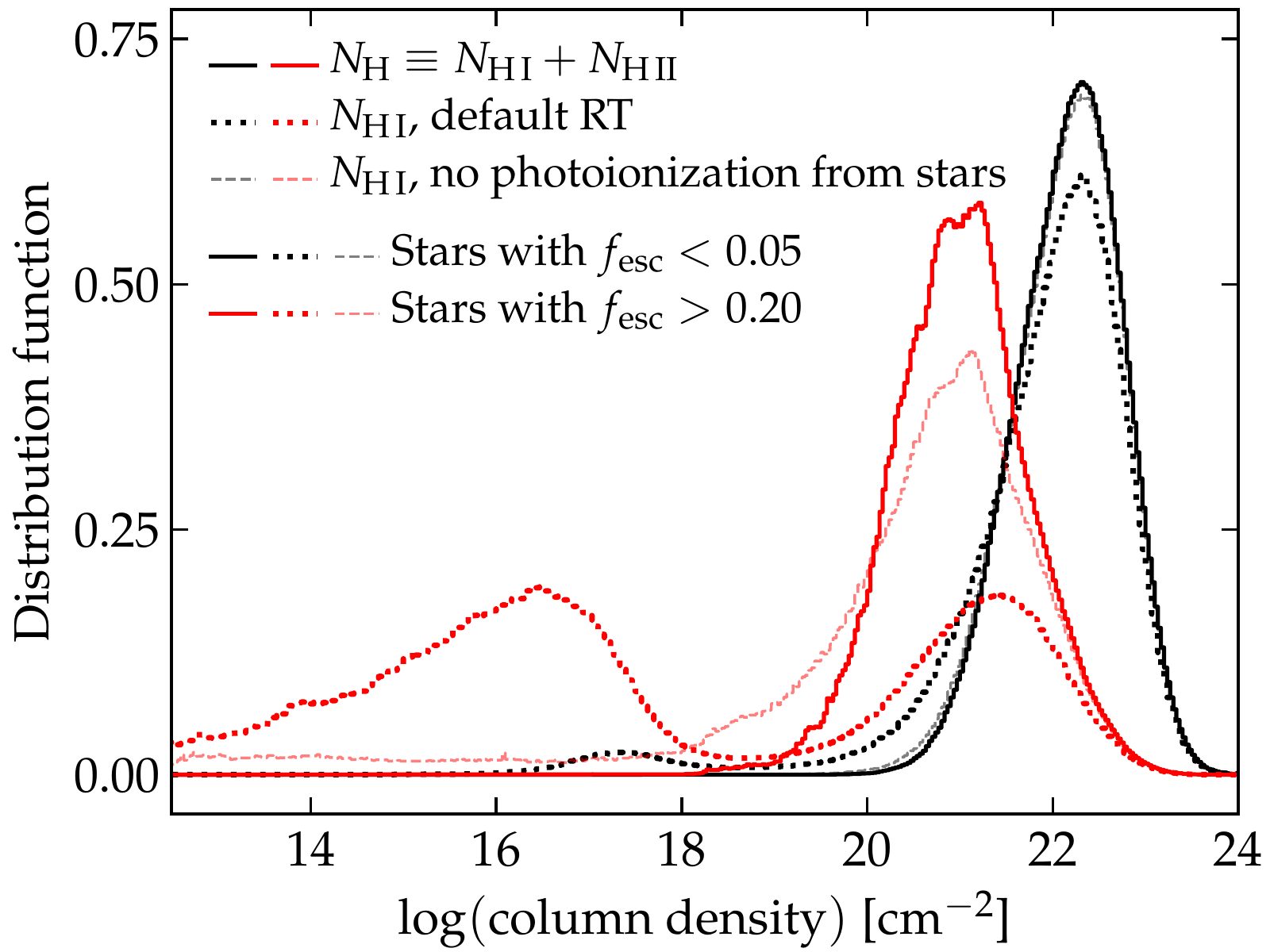}
\caption{The distribution of gas column densities from stars $\lesssim10$\,Myr old in galaxies around $\log\Ms\sim8$ out to the halo virial radius. We include 100 random sightlines for each star particle. All stars are weighted equally. Only simulations at $m_b\sim900\,\Msun$ mass resolution are used in this analysis. The black and red colors represent stars with $\fesc<0.05$ and $\fesc>0.2$, respectively. The solid lines show the distribution of $N_{\rm H}$ (neutral and ionized). The high-$\fesc$ stars tend to locate in regions surrounded by lower-$N_{\rm H}$ sightlines, likely cleared by feedback, than the low-$\fesc$ stars. The dotted lines present the distribution of $N_{\rm H\,I}$, where we use the ionization states determined from our default MCRT calculations. As a proof of concept, the thin dashed lines show the distribution of $N_{\rm H\,I}$ without photoionization from stars. The high-$\fesc$ stars are surrounded by a large fraction of optically-thin sightlines that must be ionized collectively by the young stars in the galaxy. However, the low-$\fesc$ stars are almost fully covered by optically-thick sightlines.}
\label{fig:col}
\end{figure}

For a given star particle in our simulation, we can use the octree to calculate the hydrogen column density from the star particle to the virial radius of the halo along a given sightline, in which we use the ionization states determined by the MCRT code to compute the column density of neutral hydrogen. In Fig. \ref{fig:col}, we compare the column density distribution for selected stars younger than 10\,Myr in galaxies around $\log\Ms\sim8$ ($\pm0.25$\,dex; only simulations run at $900\,\Msun$ are included), where the sample average $\fave$ peaks at 0.2 (see Fig. \ref{fig:fescM}). The black and red lines show stars with {\em individual-star} escape fraction $\fesc<0.05$ and $\fesc>0.2$, respectively. From every star particle, we compute the column density out to the virial radius along 100 random directions. Each stars is weighted equally when calculating the distribution function.

The solid lines show the distribution function of total (neutral and ionized) hydrogen column density ($N_{\rm H}=N_{\rm H\,I}+N_{\rm H\,II}$), while the dotted lines show that only for neutral hydrogen ($N_{\rm H\,I}$), with ionization states taken from our default MCRT calculations. As a proof of concept, we also redo the MCRT calculations without photoionization from stars (but including the uniform ionizing background and collisional ionization) and show the resulting distribution of $N_{\rm H\,I}$ in Fig. \ref{fig:col} with the thin dashed lines. Our results highlight two physical processes that are crucial to the escape of ionizing photons. First of all, comparing the black and red solid lines, we find that stars leaking ionizing photons effectively (e.g. $\fesc>0.2$; red) tend to locate in regions with lower $N_{\rm H}$ around compared to stars that have much lower $\fesc$ (black). These regions are presumably cleared by stellar feedback (e.g. SN bubbles; see Fig. \ref{fig:leak} and Section \ref{sec:geo}). Second, we \linebreak find a large fraction of optically-thin ($N_{\rm H\,I}\lesssim2\times10^{17}\,\cm^{-2}$) sightlines surrounding stars with high $\fesc$, through which ionizing photons can escape freely. More importantly, comparing the red dashed and dotted lines, we argue that these optically-thin channels around young stars must be self-ionized by these stars. This is more likely to happen in regions where a large number of stars have formed in the past 10\,Myr. In contrast, stars that have much lower $\fesc$ tend to be fully embedded in optically-thick ($N_{\rm H\,I}\gg2\times10^{17}\,\cm^{-2}$) sightlines. These stars are not sufficient to highly ionize the surrounding gas, making it difficult for their ionizing photons to escape.

Although we only show the column density distribution for all stars younger than 10\,Myr in galaxies around $\log\Ms\sim8$ in Fig. \ref{fig:col}, we have confirmed that all our conclusions still hold if we compare stars in a narrow age bin or in galaxies at a different mass.

\begin{figure}
\centering
\includegraphics[width=\linewidth]{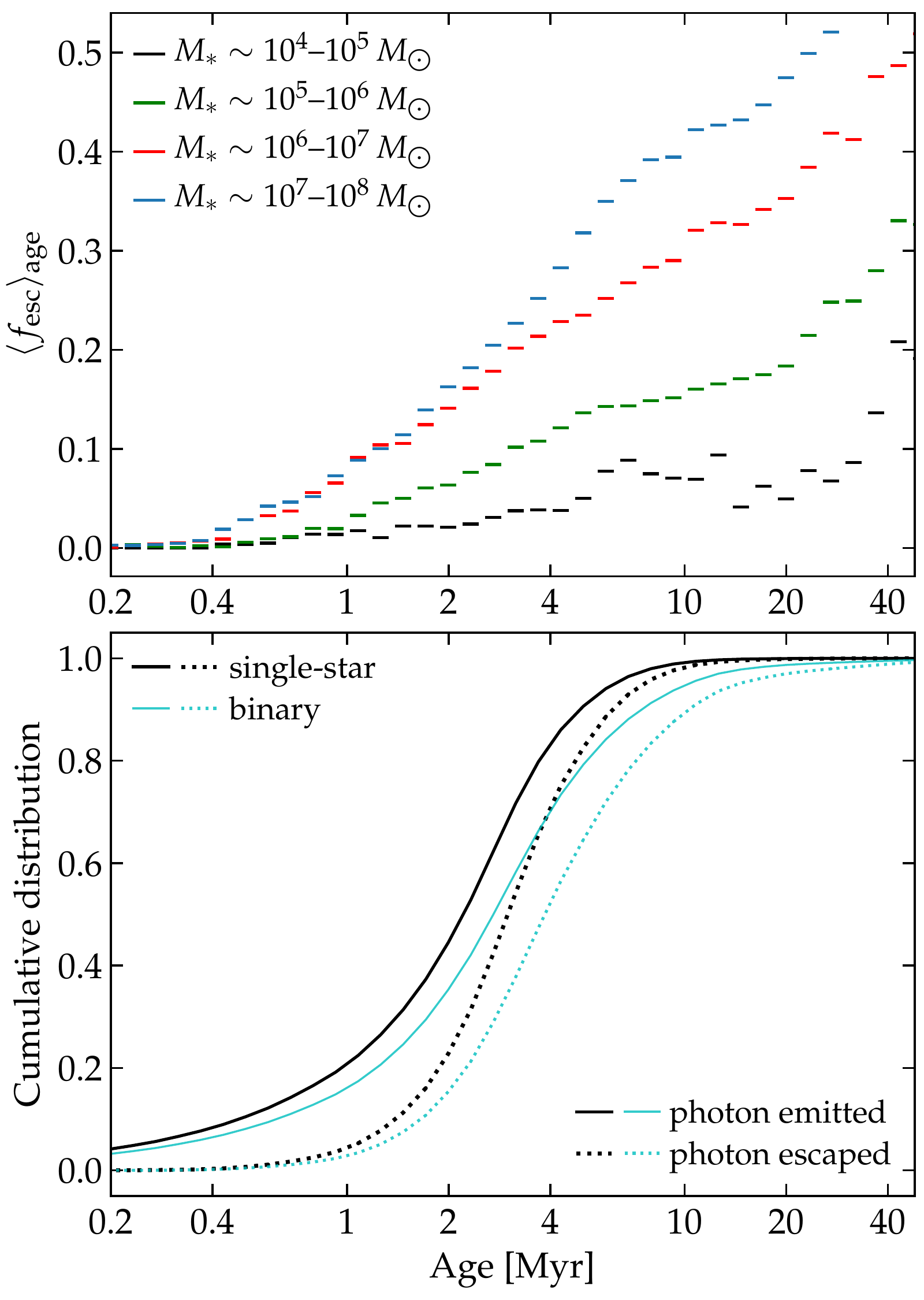}
\caption{{\em Top}: The average escape fraction over individual stars in galaxies in four stellar mass bins for every $1/15$ dex in logarithmic age. Only simula- tions at $900\,\Msun$ and better resolution are included. $\fage$ increases with stellar age for galaxies of all masses. $\fage$ increases with stellar mass at all ages, in line with the increase of sample average $\fave$ with stellar mass in this mass range. {\em Bottom}: The cumulative distribution of ionizing photons emitted (solid) and escaped (dashed) as a function of stellar age. In single-star case (black), 50\% of the escaped ionizing photons come from stars 1-- 3\,Myr old, while the rest 50\% from stars 3--10\,Myr old. The binary models extend the distribution to slightly later times.}
\label{fig:fage} 
\end{figure}

\begin{figure}
\centering
\includegraphics[width=\linewidth]{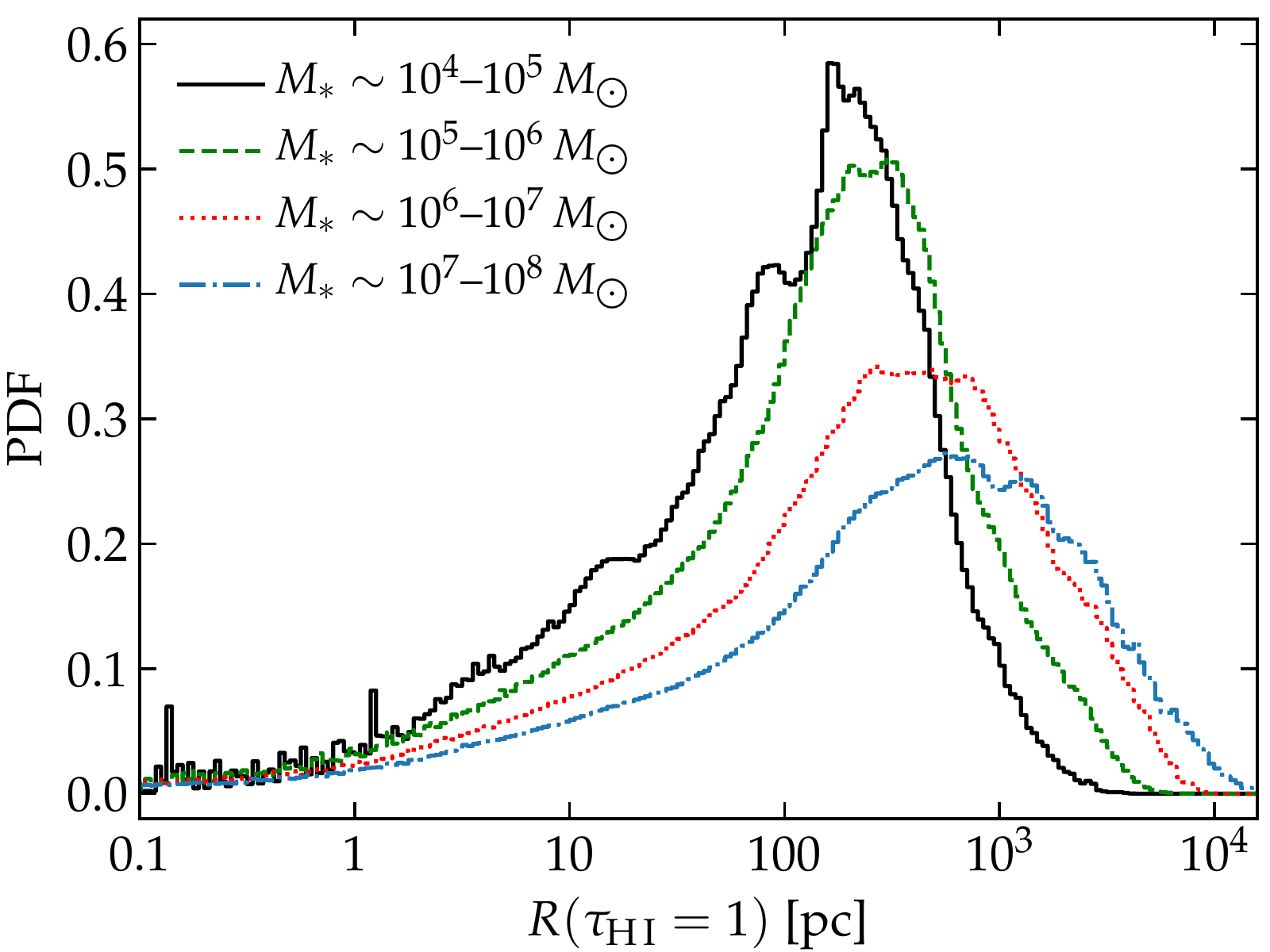}
\caption{The distribution of radius from stars younger than 10\,Myr out to which the optical depth of neutral hydrogen $\tau_{\rm H\,I}=1$ at the Lyman limit (a proxy to the distance an ionizing photon may travel) for galaxies in the four stellar mass bins. We include 100 random sightlines for each star particle and all stars are weighted equally. Only simulations at $900\,\Msun$ resolution or better are included. In low-mass galaxies, about 60\% of the sightlines from young stars become optically thick within 100\,pc. The fraction of sightlines with $R(\tau_{\rm H\,I}=1)\gtrsim100\,\pc$ and the median $R(\tau_{\rm H\,I}=1)$ increase with stellar mass, meaning a larger fraction of the ionizing photons can travel to longer distances in relatively high-mass galaxies. This indicates that feedback cannot clear the surroundings of young stars in low-mass galaxies.}
\label{fig:absdist} 
\end{figure}
	
\subsection{Escape fraction by stellar age}
\label{sec:age}
We define the average of escape fraction over individual {\em stars} in a narrow age bin, $\fage$, for a selected population of galaxies from our simulations. In the top panel of Fig. \ref{fig:fage}, we present $\fage$ as a function of stellar age for galaxies in four stellar mass bins, where we calculate $\fage$ for every $\frac{1}{15}$\,dex in logarithmic age. Note that we only use simulations at $900\,\Msun$ or better resolution and single-star models for post-processing calculations. 

We find $\fage$ increases with age for galaxies of all masses. A large fraction of the young stars are still embedded in their birth clouds, so they tend to have low $\fesc$ on average. As feedback from these stars starts to destroy the birth clouds, blow out superbubbles in the ISM, and clear low-column-density sightlines, their ionizing photons can escape more easily, thereby increasing $\fage$ at later times. At a given age, $\fage$ tends to increase with stellar mass, in line with the trend between the sample average $\fave$ and $\Ms$ in Fig. \ref{fig:fescM}. In more massive galaxies, stars 10\,Myr old have an average $\fage\sim0.4$; even stars 1--3\,Myr old on average leak 10--20\% of their ionizing photons, most of which are likely from stars formed in an accelerated shell at the edge of a superbubble (e.g. Fig. \ref{fig:leak}). In galaxies under $\Ms\sim10^5\,\Msun$, however, almost no ionizing photon from stars younger than 3\,Myr are able to escape; stars 10\,Myr old only have $\fage$ lower than 0.1. After 20\,Myr, $\fage$ increases more significantly with age, indicating that this is the time-scale on which feedback eventually clears some sightlines in such low-mass galaxies, but stars older than 20\,Myr no longer have a high ionizing photon production efficiency.

For completeness, we show the cumulative distribution of ionizing photons emitted (solid) and escaped (dashed) as a function of stellar age in the bottom panel of Fig. \ref{fig:fage}. The distribution functions are calculated using all galaxies in our simulations at $900\,\Msun$ resolution or better. We find nearly identical distributions for galaxies in a narrow mass bin or at $7000\,\Msun$ resolution. When using single-star models (black), we find nearly 80\% of the ionizing photons are emitted from stars younger than 3\,Myr and the rest 20\% from stars 3--10\,Myr old, as the ionizing photon emissivity decreases dramatically after 3\,Myr following the death of the most massive stars. As $\fage$ is low in the first Myr and increases with age, we find that nearly 50\% of the escaped ionizing photons are from stars 1--3\,Myr old and the other 50\% from stars 3--10\,Myr old. Alternatively, 90\% of the escaped photons are from stars 1--5\,Myr old.

In Section \ref{sec:binary}, we mention that binary models produce more ionizing photons after 3\,Myr than single-star models owing to mass transfer and stellar mergers. These extra photons tend to escape efficiently given the relatively high $\fage$ after 3\,Myr. When using binary models (cyan), we find 55\% (35\%) of the emitted (escaped) ionizing photons come from stars younger than 3\,Myr, 40\% (55\%) from stars 3--10\,Myr old, and the rest 5\% (10\%) from stars over 10 Myr old. We find binary stars only increase the number of ionizing photons escaped by 60--80\%, much lower than that reported in previous works \citep[cf. a factor of 3 or more; e.g.][]{Ma:2016,Rosdahl:2018}. This is likely due to the fact that a large fraction of the escaped ionizing photons are from stars younger than 3\,Myr (when binary evolution is subdominant) in our simulations, thus reducing the relative effects of the extra photons from binaries after 3\,Myr.

\subsection{Why does $\fave$ decrease at the low-mass end?}
\label{sec:col}
In Section \ref{sec:fM} and Fig. \ref{fig:fescM}, we show that the sample average $\fave$ decreases with decreasing stellar mass below $\Ms\sim10^8\,\Msun$. There is a similar trend with halo mass at $\Mvir\lesssim10^{11}\,\Msun$ if dust attenuation is ignored (e.g. the left column in Fig. \ref{fig:fMcomp}). We study why $\fave$ decreases at the low-mass in this section. 

From each star particle younger than 10\,Myr, we calculate the radius out to which the optically depth of H\,{\sc i} $\tau_{\rm H\,I}=1$ at the Lyman limit from the particle along 100 random sightlines, using a similar method to what described in Section \ref{sec:feedback}. This radius is an approximate measure to the distance an ionizing photon may travel before absorbed by neutral hydrogen. In Fig. \ref{fig:absdist}, we show the distribution of this radius for galaxies in four stellar mass bins from $\Ms\sim10^4$--$10^8\,\Msun$. Only simulations at $900\,\Msun$ resolution or better are used. We weight all stars equally in the distribution functions. If a sightline has $\tau_{\rm H\,I}<1$ out to the virial radius, we set this radius to infinity \\ (not shown in Fig. \ref{fig:absdist}, the fraction of such optically-thin sightlines increases with $\Ms$). In low-mass galaxies ($\Ms\lesssim10^5\,\Msun$), 60\% of the sightlines from stars younger than 10\,Myr turn optically thick in 100\,pc. The fraction of sightlines with $R(\tau_{\rm H\,I}=1)\gtrsim100\,\pc$, as well as the median $R(\tau_{\rm H\,I}=1)$, increases with stellar mass. The results in Fig. \ref{fig:absdist} suggest that most ionizing photons in low-mass galaxies are absorbed in a short range, while a larger fraction of the ionizing photons can reach larger distances (from a few 100\,pc to 10\,kpc) in relatively high-mass galaxies.

\begin{figure}
\centering
\includegraphics[width=\linewidth]{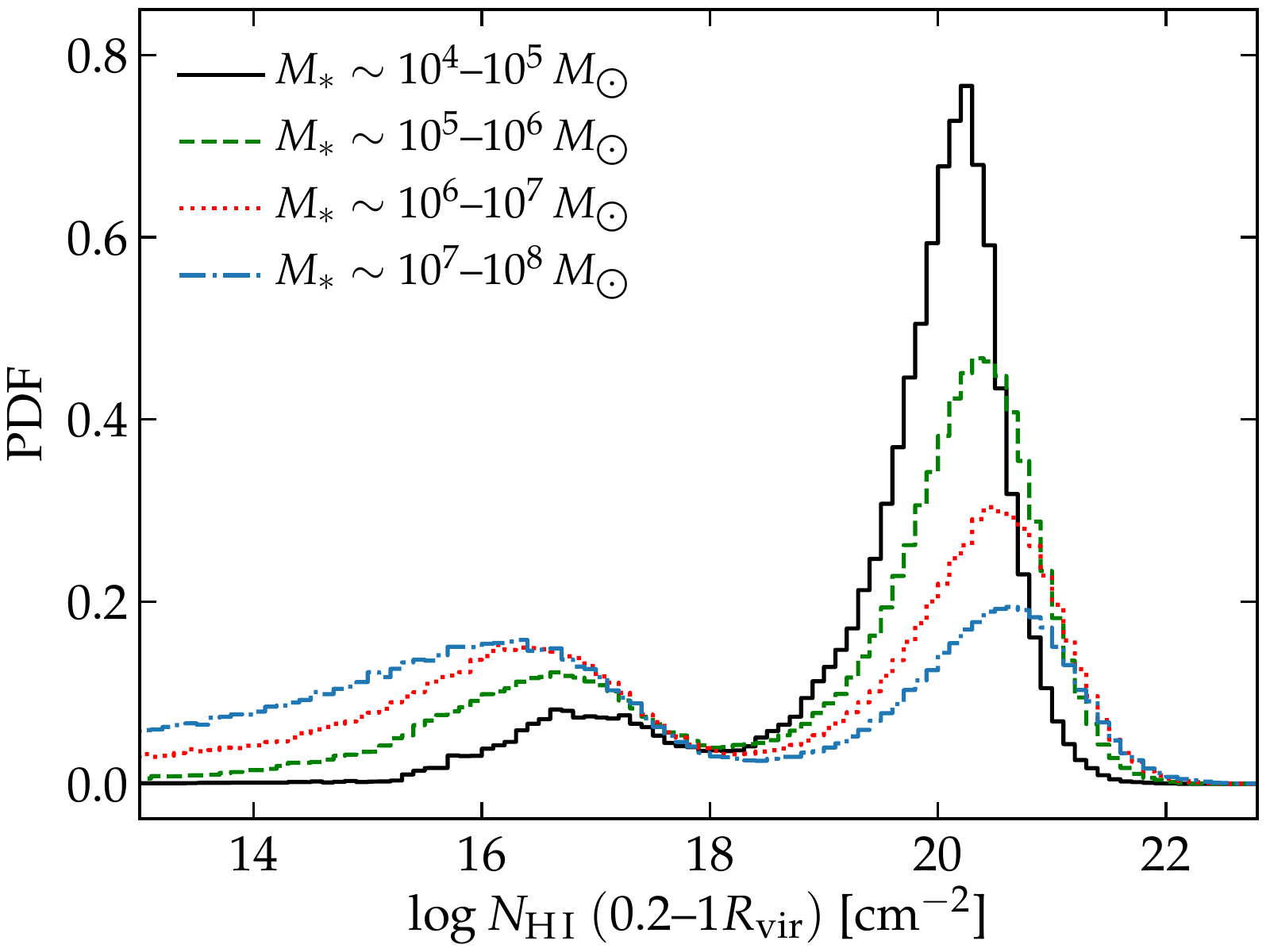}
\caption{The distribution of $N_{\rm H\,I}$ integrated radially from 0.2--$1\Rvir$ in the halo of galaxies in four stellar mass bins. We include 1000 radial directions for each galaxy. All galaxies are equally weighted. The fraction of optically-thin sightlines increases with stellar mass. The low-mass galaxies are nearly fully surrounded by optically-thick gas in the halo.}
\label{fig:halocol} 
\end{figure}

In Fig. \ref{fig:halocol}, we show the distribution of $N_{\rm H\,I}$ integrated radially from 0.2--$1\Rvir$ for galaxies in the same stellar mass bins, where we calculate $N_{\rm H\,I}$ along 1000 directions for each galaxy. We weight all galaxies equally. The distribution of $N_{\rm H\,I}$ from the outer ISM to the halo is bimodal, similar to the red dotted line in Fig. \ref{fig:col}. The fraction of optically-thin sightlines ($N_{\rm H\,I}\lesssim2\times10^{17}\,\cm^{-2}$) decreases significantly with decreasing stellar mass. This suggests a high covering fraction of optically-thick neutral gas ($N_{\rm H\,I}\gtrsim2\times10^{17}\,\cm^{-2}$) in the halo around low-mass galaxies.\footnote{We have checked that this trend below $\Ms\sim10^8\,\Msun$ still holds for simulations run at $\mb\sim7000\,\Msun$ resolution. However, we find that the distribution of $N_{\rm H\,I}$ becomes independent of stellar mass at $\Ms\gtrsim10^8\,\Msun$. This indicates that the nearly constant $\fave$ above $\Ms\sim10^8\,\Msun$ without dust attenuation (bottom left panel in Fig. \ref{fig:fMcomp}) is likely due to a constant opening angle of optically-thin sightlines at the high-mass end.} We reiterate that the optically-thin paths are photoionized by the young stars collectively in the galaxy (cf. Fig. \ref{fig:col} and Section \ref{sec:feedback}), presumably through some low-column-density paths channels pre-cleared by feedback.

In \citet{Ma:2018a}, we show the stellar mass and average SFR scale with halo mass as $\propto \Mvir^{1.5}$, which means star formation is less efficient in low-mass galaxies than in their high-mass counterparts. Moreover, the mass-weighted gas temperature in the halo (roughly independent of radius in 0.2--$1\Rvir$) decreases from $10^5$\,K for galaxies in $\Ms\sim10^7$--$10^8\,\Msun$ to $10^4$\,K for those in $\Ms\sim10^4$--$10^5\,\Msun$, making collisional ionization less effective in the halo of low-mass galaxies. Combined the results from Figs. \ref{fig:fage}--\ref{fig:halocol}, we suggest that the low $\fave$ in low-mass galaxies owes to a combination of reasons as follows. First, feedback is not strong enough to blow out kpc-scale superbubbles around stars $\lesssim10$\,Myr old and trigger star formation in the dense shell surrounding the bubble simultaneously. This can be seen from the fact that only a small fraction of the ionizing photons from young stars can travel more than 100\,pc before absorbed in low-mass galaxies (Fig. \ref{fig:absdist}). Second, the young stars cannot fully ionize a large number of channels throughout the halo (Fig. \ref{fig:halocol}), so stars of all ages tend to have low escape fractions on average in the  low-mass galaxies (the top panel in Fig. \ref{fig:fage}). Both arguments above are likely resulted from the low star formation efficiencies in these galaxies, namely low-mass galaxies do not form sufficient stars coherently to clear some low-column-density paths and to fully ionize \linebreak these sightlines. Finally, the low gas temperatures in the halo make the neutral gas covering fraction higher around low-mass galaxies, at least partly responsible to the low $\fage$ in Fig. \ref{fig:fage}.

\section{Discussion}
\label{sec:discussion}
\subsection{The impact of sub-grid recipes}
\label{sec:subgrid}
In Section \ref{sec:intro}, we mentioned that the prediction of $\fesc$ from hydrodynamic simulations of galaxy formation might be sensitive to the `sub-grid' models implemented in these simulations. In particular, \citet{Ma:2015} found $\fave\lesssim0.05$ using a sample of three simulations spanning $\Mvir\sim10^9$--$10^{11}\,\Msun$ at $z\geq5$ run with the FIRE-1 version of {\sc gizmo} \citep[see][for details]{Hopkins:2014}, whereas in this paper, we find $\fave\sim0.2$ in $\Mvir\sim10^{9.5}$--$10^{11}\,\Msun$ in FIRE-2 simulations, both using single-star stellar population models.

To test possible subtle effects of different sub-grid treatments, we re-run simulation z5m11b (with halo mass $\Mvir\sim4\times10^{10}\,\Msun$ at $z=5$) from the initial condition to $z=5$ for more than 20 times with almost all possible combinations of choices for hydrodynamic method, sub-grid models, etc., as listed below.
\begin{itemize}
\item Hydrodynamic solver: P-SPH \citep[][used for FIRE-1]{Hopkins:2013a} and MFM \citep[][FIRE-2]{Hopkins:2015}.
\item Density threshold for star formation in $\nc=50$--$1000\,\cm^{-3}$. The default value for FIRE-1 (FIRE-2) is 50 (1000) cm$^{-3}$. We do not experiment with higher $\nc$ given our resolution.
\item The non-conservative FIRE-1 and the more accurate, conservative FIRE-2 SN coupling algorithms \citep[see][for detailed descriptions and comparisons]{Hopkins:2018a}.
\item The default self-gravitating criteria for star formation (used for both FIRE-1 and FIRE-2) and the stricter version from \citet{Grudic:2018} (see also \citealt{Ma:2020a}).
\item The star formation efficiency per local free-fall time $\epsilon\sim0.1$--1 (default is 1 in both FIRE-1 and FIRE-2).
\item The maximum search radius for gas particles for SN coupling in 0.2--10\,kpc (default is 2\,kpc in both FIRE-1 and FIRE-2).
\item More subtle changes from FIRE-1 to FIRE-2 including the inclusion of an artificial pressure floor from \citet{Truelove:1997} for the P-SPH method, cooling functions, and recombination rates.
\end{itemize}

All the tests we run produce statistically indistinguishable star formation histories for galaxy z5m10b. We also run post-processing calculations on all these simulations using our MCRT code to calculate $\fesc$. We compare the time average of $\fesc$ over 48 snapshots in $z=5$--10 for this galaxy, $\fave_t$. We find that runs using the P-SPH method generally produce $\fave_t\sim0.1$, while those using the MFM method predict $\fave_t\sim0.2$. The star formation criteria, SN coupling algorithms, etc., make more subtle differences. Note that the P-SPH method smooths the density field using a quintic spline kernel over 64 nearest particles, thereby lowering the effective hydrodynamic resolution compared to the MFM method at the same particle mass. The difference in the hydrodynamic solver is in line with the difference in Fig. \ref{fig:fescM} between simulations at $7000\,\Msun$ and \linebreak those at $900\,\Msun$ or better resolution, as some marginally optically-thin channels with $\tau_{\rm H\,I}\lesssim1$ from which ionizing photons are able to \linebreak escape would be under-resolved, or over-smoothed with optically-thick sightlines nearby and become $\tau_{\rm H\,I}\gtrsim1$ at lower resolution.

However, we are not able to isolate the reason that causes the difference in $\fesc$ between FIRE-1 and FIRE-2, presumably because of the complex, non-linear nature this problem.

In all the test runs, we identify the same configuration as those shown in Fig. \ref{fig:leak} for vigorously star-forming regions where ionizing photons leak efficiently, namely a feedback-driven (sub-)kpc-scale superbubble surrounded by an accelerated, star-forming shell. New stars formed in the shell present a clear age gradient at the edge of the bubble, among which the relatively older ones, despite younger than 3\,Myr, are already inside the low-density region. This happens in all sub-grid models of star formation and SN feedback we have tested \citep[see also][]{Yu:2020}. We caution that our simulations do not fully resolve the radiative shock at the front of the superbubble nor include a chemical network for dust and $\rm H_2$ molecule for more \linebreak accurate star formation prescriptions, so we might not produce the correct amount of stars formed in the shell or the exact time-scale for star formation to happen there. However, we emphasize the fact that similar phenomenon has been observed as supergiant shells in the Large Magellanic Clouds (e.g. \citealt{Pellegrini:2012,Dawson:2013}; see also supershells and propagated star formation, e.g. \citealt{Heiles:1979,McCray:1987}). 

\subsection{Which galaxies provide the most ionizing photons?}
\label{sec:budget}
In Section \ref{sec:fM}, we find $\fave$ increases with halo mass in $\Mvir\sim10^8$--$10^{9.5}\,\Msun$, turns nearly constant in $\Mvir\sim10^{9.5}$--$10^{11}\,\Msun$, and decreases with halo mass above $\Mvir\sim10^{11}\,\Msun$. In the literature, the dependence of $\fave$ on $\Mvir$ has been studied in state-of-the-art simulations with sophisticated chemical network and/or on-the-fly radiation-hydrodynamics. These simulations are fairly expensive so \linebreak some of them are run in small cosmological volumes (box size less than $\sim10\,{\rm cMpc}$) and/or stopped at relatively high redshifts ($z\gtrsim8$) (see e.g. \citealt{Wise:2014,OShea:2015,Paardekooper:2015,Xu:2016b}). These studies found that $\fave$ decrease with \linebreak halo mass, from $\fave\sim0.5$ at $\Mvir\lesssim10^7\,\Msun$ to $\fave\lesssim0.05$ at $\Mvir\gtrsim10^8\,\Msun$ \citep[e.g.][]{Wise:2014,Xu:2016b}. Intriguingly, \citet{Xu:2016b} found that $\fave$ starts to increase at $\Mvir\gtrsim10^9\,\Msun$ to $\fave\sim0.1$--0.2 at $\Mvir\sim10^{9.5}\,\Msun$, in line with what we find in our simulations. Nonetheless, the simulation from \citet{Xu:2016b} only contains a small number of halos at $\Mvir\sim10^{9.5}\,\Msun$ while no \linebreak halo at higher masses. Our results thus complement these previous studies by extending the $\fave$--$\Mvir$ relation to the massive end.

In our simulations, the average SFR, and thereby the ionizing photon emissivity $\langle Q_{\rm ion} \rangle$, scale with halo mass as $\propto\Mvir^{1.5}$. A nearly constant $\fave$ gives $\langle Q_{\rm esc} \rangle \propto\Mvir^{1.5}$ for the rates of escaped photons at intermediate halo mass from $\Mvir\sim10^9$--$10^{11}\,\Msun$, which is also confirmed directly in our post-processing calculations. The scaling becomes steeper (shallower) as $\langle Q_{\rm esc} \rangle \propto \Mvir^{2.5}$ ($\propto \Mvir^{1.0}$) at the low- (high-)mass end as $\fave$ decreases. At a given halo mass, both the SFR (and hence $\langle Q_{\rm ion} \rangle$) and $\fave$ increase with redshift (see fig. 7 in \citealt{Ma:2018a} and Fig. \ref{fig:fMcomp}), so $\langle Q_{\rm esc} \rangle$ also increases with redshift. The best-fit normalization of the broken power-law function to our data is $\langle Q_{\rm esc} \rangle\sim1$--$4\times10^{53}\,\s^{-1}$ from $z=6$--10 at $\Mvir=10^{11}\,\Msun$, where we use single-star stellar population models. By convolving the broken power-law function of $\langle Q_{\rm esc} \rangle$--$\Mvir$ relation with the halo mass functions \citep[HMFs;][]{Murray:2013} at $z\geq5$, we obtain the number density of ionizing photons escaped to the IGM as $\dot{n}_{\rm ion}\sim10^{51.2}$--$10^{50.6}\,\s^{-1}\,\Mpc^{-3}$, decreasing with redshift from $z=6$ to 10. Binary stars will enhance $\dot{n}_{\rm ion}$ by about $\sim60$--80\% (Section \ref{sec:binary}). Our estimate of $\dot{n}_{\rm ion}$ is in broad agreement with what derived from the most recent constraints on the reionization history \citep[e.g.][]{Mason:2019a}. Here we present the quantitative details on $\langle Q_{\rm esc} \rangle$ and {\linebreak} $\dot{n}_{\rm ion}$ only for completeness. Given the non-convergence in our simulations, we emphasize that these numbers likely suffer a factor of 2 uncertainties. They should be used with caution.

Now we consider the distribution of $\dot{n}_{\rm ion}$ per logarithmic halo mass, $\dd \dot{n}_{\rm ion}/\dd \log\Mvir$. The HMF can be well described by a power-law function at the low-mass end and an exponential function at the high-mass end, $\dd n/\dd \log\Mvir \sim \Mvir^{1-\alpha} \exp(-\Mvir/\Mvir^{\ast})$, where $\Mvir^{\ast}$ is some characteristic mass \citep{Schechter:1976}. As $\dd \dot{n}_{\rm ion}/\dd \log\Mvir = \langle Q_{\rm esc} \rangle \, \dd n/\dd \log\Mvir$, for the canonical $\alpha=2$ slope, $\dd \dot{n}_{\rm ion}/\dd \log\Mvir$ increases with $\Mvir$ at the low-mass end and decreases dramatically above $\Mvir^{\ast}$. We find at $z\sim6$, $\dd \dot{n}_{\rm ion}/\dd \log\Mvir$ peaks at approximate $\Mvir\sim10^{10.5}\,\Msun$ ($\Ms\sim10^8\,\Msun$), which means that intermediate-mass galaxies dominate the cosmic ionizing photon budget at $z\sim6$ \citep[see also][]{Naidu:2020}. However, the HMF starts to decline at a much smaller mass at $z\sim10$, so we find $\dd \dot{n}_{\rm ion}/\dd \log\Mvir$ peaks at $\Mvir\sim10^{9}\,\Msun$ ($\Ms\lesssim10^6\,\Msun$). Our results suggest that low-mass galaxies dominate the ionizing photon budget at early times, while more massive galaxies take over near the end of reionization\footnote{\citet{Faucher-Giguere:2020} discussed the apparent tension at $z\sim3$ between the integrated constraint of $\fesc\sim0.01$--0.02 from Lyman-$\alpha$ forest and the observed $\fesc\sim0.1$ for luminous $z\sim3$ galaxies from \citet{Steidel:2018}. If the trend from our simulations continues to $z\sim3$, the tension described above may be alleviated, as faint galaxies do not contribute significant ionizing photons that escape to the IGM.} \citep[e.g.][]{Finkelstein:2012,Finkelstein:2019,Robertson:2013,Robertson:2015,Faucher-Giguere:2020,Naidu:2020,Yung:2020}.

\section{Conclusions}
\label{sec:conclusion}
In this paper, we use a sample of 34 high-resolution cosmological zoom-in simulations of $z\geq5$ galaxies run with the FIRE-2 version of the source code {\sc gizmo} and explicit models for the multi-phase ISM, star formation, and stellar feedback in \citet{Hopkins:2018b}. Our sample consists of simulations run at baryonic mass resolution $m_b\sim7000\,\Msun$, $900\,\Msun$, and $100\,\Msun$. We post-process over 8500 relatively well-resolved galaxy snapshots from all zoom-in regions with a Monte Carlo radiative transfer code for ionizing radiation to calculate $\fesc$ and the gas ionization states. Our default calculations assume a constant dust-to-metal ratio of 0.4 in gas below $10^6$\,K (no dust at higher temperatures) and a SMC-like extinction curve from \citet{Weingartner:2001}. We consider both the single-star and binary models from the BPASS stellar population synthesis models to calculate the ionizing photon emissivity for every star particle in our simulations \citep[v2.2.1;][]{Eldridge:2017}.

We study the sample average $\fave$ (i.e. the average of instantaneous $\fesc$ over all galaxies at all redshifts for a given halo/stellar mass bin) and its dependence on halo or stellar mass, redshift, dust, and stellar population models. We also explore the key physics that governs the escape of ionizing photons in our simulations. 

Our main findings include the following.

(i) Both the instantaneous $\fesc$ and SFR exhibit strong variability on short time-scales (Section \ref{sec:fins}, Fig. \ref{fig:sfr}). There is usually a time delay between the rising of $\fesc$ and the rising of SFR at the beginning of a starburst, because it takes some time for feedback to clear the sightlines for ionizing photons to escape. A galaxy may have a high $\fesc$ but low SFR at some epochs, meaning that it is not leaking a large number of ionizing photons. The instantaneous $\fesc$--$\Mvir$ relation shows enormous scatter, with $\fesc$ ranging from $\lesssim10^{-4}$ to 1 at fixed $\Mvir$ (Fig.~\ref{fig:fesc}).

(ii) Our results on the sample average $\fave$ do not fully converge with resolution. Simulations run at $7000\,\Msun$ resolution tend to produce systematically lower $\fave$ than those run at $900\,\Msun$ or better resolution. Simulations with 900 and $100\,\Msun$ resolution produce consistent results on $\fave$. Nonetheless, the qualitative trends in the $\fave$--$\Mvir$ ($\Ms$) relation are robust (Section \ref{sec:fM}, Fig. \ref{fig:fescM}).

(iii) In our default dust model, $\fave$ increases with halo mass in $\Mvir\sim10^8$--$10^{9.5}\,\Msun$, becomes roughly constant in $\Mvir\sim10^{9.5}$--$10^{11}\,\Msun$, and decreases at $\Mvir\gtrsim10^{11}\,\Msun$ (left, Fig. \ref{fig:fescM}). $\fave$ also increases with stellar mass in $\Ms\sim10^4$-- $10^8\,\Msun$ and decreases at $\Ms\gtrsim10^8\,\Msun$ (right, Fig. \ref{fig:fescM}). The declining $\fave$ at the high-mass end is due to dust attenuation (Section \ref{sec:dust}; left column, Fig. \ref{fig:fMcomp}).

(iv) For single-star models, $\fave\sim0.2$ around $\Ms\sim10^8\,\Msun$ and $\Mvir\sim10^{10.5}\,\Msun$ (for simulations at $900\,\Msun$ or better resolution, $\fave\sim0.1$ for those at $7000\,\Msun$ resolution). The binary stars boost $\fave$ by 25--35\%, the ionizing photon emissivity by 20--30\%, and therefore the number of photons {\em escaped} by 60--80\% (Section \ref{sec:binary}; middle column, Fig. \ref{fig:fMcomp}). The effect of binary stars is modest, \linebreak as a considerable fraction of stars younger than 3\,Myr leak ionizing photons efficiently (see below).

(v) Galaxies at $z\geq8$ tend to have systematically higher $\fave$ than those at $z<8$, suggesting a decreasing $\fesc$ toward lower redshift (right column, Fig. \ref{fig:fMcomp}).

(vi) We find a common geometry for vigorously star-forming regions that leak ionizing photons efficiently. They usually contain (or a part of) a feedback-driven, kpc-scale superbubble surrounded by a dense, star-forming shell. The shell is also accelerated, leaving an age gradient in the newly formed stars at the bubble edge. Stars formed slightly earlier in the shell, despite younger than 3\,Myr, are already at the inner side of the shell. These young stars, along with stars 3--10\,Myr old in the bubble, can fully ionize the low-column-density sightlines surrounding the bubble, allowing a large fraction of their ionizing photons to escape (Section \ref{sec:geo}, Fig. \ref{fig:leak}). 

(vii) Young stars ($\lesssim10$\,Myr) with high $\fesc$ (measured for individual stars) preferentially locate in regions with lower column densities out to the virial radius, compared to stars with low $\fesc$. These regions are presumably cleared by stellar feedback. In addition, the low-column-density sightlines must also be ionized collectively by the young stars in these regions to become optically thin to ionizing {\linebreak} photons. In contrast, stars with low $\fesc$ are fully hidden in optically-thick sightlines (Section \ref{sec:feedback}, Fig. \ref{fig:col}). 

(viii) The average of $\fesc$ over stars in a given age, $\fage$, increases monotonically with stellar age in 0--40\,Myr, likely because the impact of feedback in clearing the sightlines gets stronger with time. At fixed age, $\fage$ decreases in galaxies with decreasing stellar mass in $\Ms\sim10^4$--$10^8\,\Msun$, in line with the $\fave$--$\Ms$ relation at the low-mass end (Section \ref{sec:age}, Fig. \ref{fig:fage}).

(ix) In single-star models, about a half of the {\em escaped} ionizing photons come from stars 1--3\,Myr old, while the rest from stars 3--10\,Myr old. The contribution from stars $\gtrsim10$\,Myr old is negligible. \linebreak In binary models, 35\%, 45\%, and 20\% of the escaped photons are from stars 1--3, 3--10, and $\gtrsim10$\,Myr old (Section \ref{sec:age}, Fig. \ref{fig:fage}).

(x) With decreasing stellar mass at $\Ms\lesssim10^8\,\Msun$, an increasing fraction of the ionizing photons are absorbed in a shorter range (Fig. \ref{fig:absdist}) and the covering fraction of optically-thick gas in the halo also increases (Fig. \ref{fig:halocol}). This suggests that the low $\fave$ at the low-mass end is due to a combination of inefficient star formation (and hence feedback) and low gas temperatures in the halo (Section \ref{sec:col}).

(xi) We estimate the escaped ionizing photon density based on \linebreak simple broken power-law fits to our simulation data, $\dot{n}_{\rm ion}\sim10^{50.6}$--$10^{51.2}\,\s^{-1}\,\Mpc^{-3}$, increasing with decreasing redshift from $z=6$ to 10. This is sufficient for cosmic reionization according to most recent constraints. We find low-mass galaxies ($\Ms\lesssim10^6\,\Msun$) dominate the cosmic ionizing photon budget at $z\sim10$, but intermediate-mass galaxies ($\Ms\sim10^8\,\Msun$) gradually take over toward the end of reionization at $z\sim6$ (Section \ref{sec:budget}).

In future work, we will carry out radiative transfer calculations on the resonance Lyman-$\alpha$ line \citep[e.g.][]{Smith:2019} and nebular lines like [O\,{\sc ii}] and [O\,{\sc iii}] \citep[e.g.][]{Arata:2020} to understand the proposed observational signatures of $\fesc$ (Section \ref{sec:intro}, and references \linebreak therein). We will also revisit the question of $\fesc$ as we keep improving our sub-grid recipes, resolution, and numerical methods.

\section*{Acknowledgement}
The simulations and post-processing calculations used in this paper were run on XSEDE computational resources (allocations TG-AST120025, TG-AST130039, TG-AST140023, TG-AST140064, and TG-AST190028). 
This work was supported in part by a Simons Investigator Award from the Simons Foundation (EQ) and by NSF grant AST-1715070.
AW was supported by NASA, through ATP grant 80NSSC18K1097 and HST grants GO-14734 and AR-15057 from STScI.
PFH was supported by an Alfred P. Sloan Research Fellowship, NASA ATP Grant NNX14AH35G, and NSF Collaborative Research Grant \#1411920 and CAREER grant \#1455342.
CAFG was supported by NSF through grants AST-1517491, AST-1715216, and CAREER award AST-1652522, by NASA through grant 17-ATP17-0067, and by a Cottrell Scholar Award from the Research Corporation for Science Advancement.
DK was supported by NSF grant AST-1715101 and the Cottrell Scholar Award from the Research Corporation for Science Advancement.

\section*{Data Availability Statement}
The data underlying this article will be shared on reasonable request to the corresponding author.

\bibliography{library}

\appendix

\section{The effect of resolution on $\fave$}
\label{app:restest}
\begin{figure}
\centering
\includegraphics[width=\linewidth]{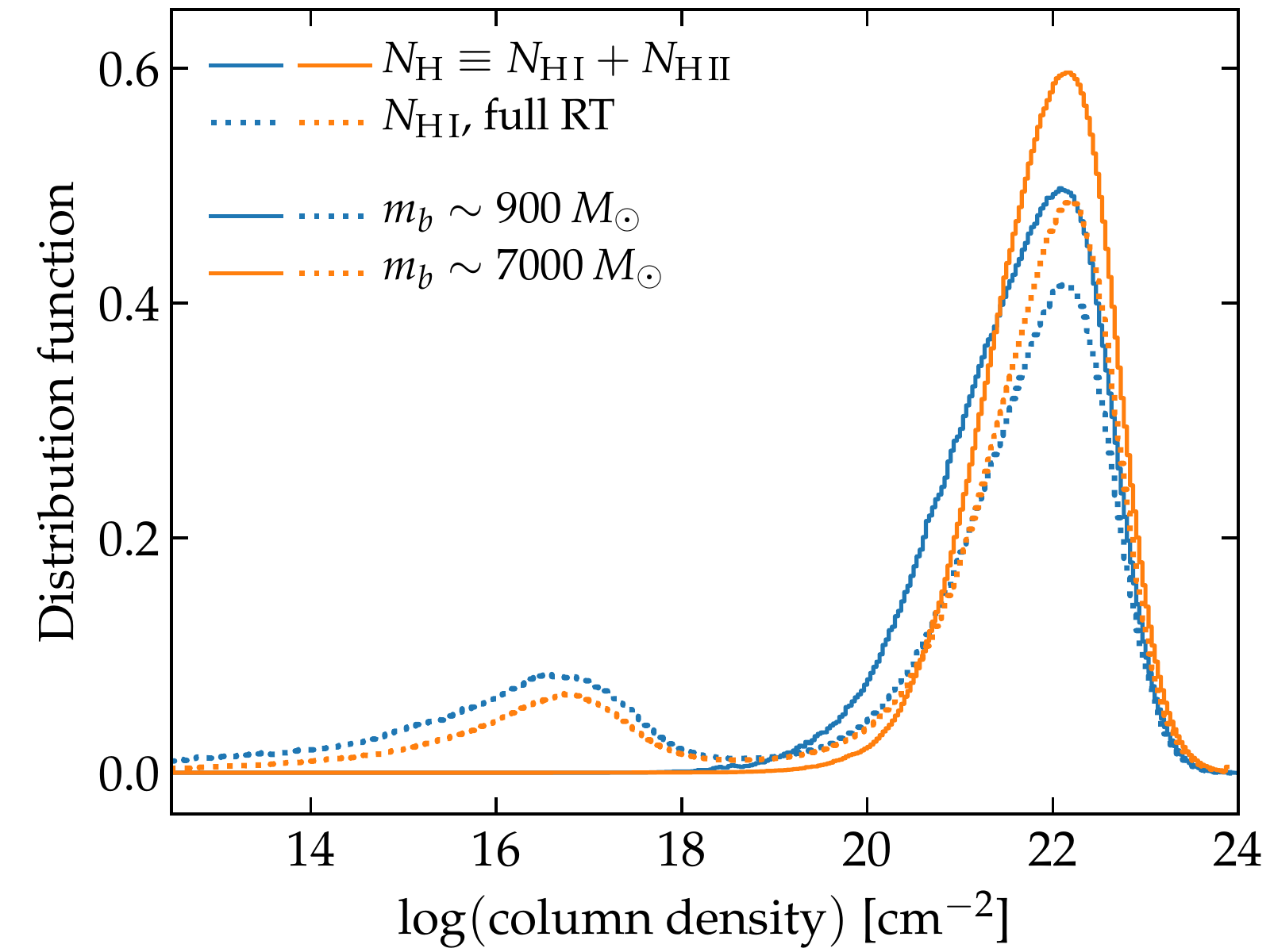}
\caption{Column density distribution from stars $\lesssim10$\,Myr old out to the halo virial radius in galaxies around $\log\Ms\sim8$. The solid and dashed lines show $N_{\rm H}$ and $N_{\rm H\,I}$, respectively. The blue and orange lines show simulations \newline at $m_b\lesssim900\,\Msun$ and $\sim7000\,\Msun$ resolution, respectively. At higher resolution, the distribution of $N_{\rm H}$ extends to lower column densities and a larger \newline fraction of sightlines are optically thin, therefore resulting in systematically higher $\fave$ at higher resolution.}
\label{fig:restest} 
\end{figure}

In Section \ref{sec:fave}, we show the sample-averaged $\fave$ does not fully converge with resolution, with simulations at $m_b\sim7000\,\Msun$ resolution producing systematically lower $\fave$ than those at $900\,\Msun$ or better resolution (Fig. \ref{fig:fescM}). We further show in Fig. \ref{fig:restest} the column density distribution from stars younger than 10\,Myr out to the halo virial radius in galaxies around $\log\Ms\sim8$. From each particle, we calculate the column densities along 100 random sightlines and the distribution functions are weighted by the ionizing photon emissivity. The solid and dashed lines show $N_{\rm H}$ and $N_{\rm H\,I}$, respectively. The colors represent simulations at $\lesssim900\,\Msun$ (blue) and $\sim7000\,\Msun$ (orange) resolution. At higher resolution, the distribution of $N_{\rm H}$ extends to lower column densities and hence a higher fraction of the sightlines become optically thin ($N_{\rm H\,I}\lesssim2\times10^{17}\,\cm^{-2}$) to ionizing \linebreak photons than at lower resolution. We speculate that the primary reason for the lower $\fave$ at lower resolution is that some of the low-column-density, optically-thin sightlines would be under-resolved, or over-smoothed and hence becomes optically thick at lower resolution, consistent with our findings in Section \ref{sec:subgrid} that SPH method tends to produce systematically lower $\fesc$ than MFM method given \linebreak the lower effective hydrodynamic resolution with SPH method. We emphasize the importance of convergence study when studying $\fesc$ in hydrodynamic simulations.

\label{lastpage}

\end{document}